\documentclass[aps, a4paper,superscriptaddress, nofootinbib, showpacs, twocolumn, 10pt]{revtex4}
\usepackage{amsmath,amssymb,bm,natbib}
\usepackage{epsfig}
\usepackage{graphicx}
\usepackage{slashed}
\newcommand{\be}{\begin{equation}}
\newcommand{\ee}{\end{equation}}
\newcommand{\bea}{\begin{eqnarray}}
\newcommand{\eea}{\end{eqnarray}}

\newcommand{\wh}{\widehat}
\newcommand{\wt}{\widetilde}
\newcommand{\nn}{\nonumber}
\begin{document}
\title{Expansion functions  in perturbative QCD and the determination of $\alpha_s(M_\tau^2)$} 
\author{Irinel Caprini}
\affiliation{Horia Hulubei National Institute for Physics and Nuclear Engineering, 
POB MG-6, 077125 Magurele, Romania}
\author{Jan Fischer}
\affiliation{Institute of Physics, Academy of Sciences of the Czech Republic, 
CZ-182 21  Prague 8, Czech Republic}
 
\begin{abstract}

The  conventional series in powers of the coupling in perturbative QCD  have zero radius of convergence and fail to reproduce the singularity of the QCD correlators like the Adler function at $\alpha_s=0$.  Using the technique of conformal mapping  of the Borel plane, combined with the "softening" of the leading singularities, we define  a set of new expansion functions that resemble the expanded correlator and share the same singularity at zero coupling. Several different conformal mappings and different ways of implementing the known  nature of the first branch-points of the Adler function in the Borel plane are investigated, in both the contour-improved (CI) and fixed-order (FO) versions of renormalization-group resummation. We prove the remarkable convergence properties of a set of new CI expansions and use them
for a determination of the strong coupling from the hadronic $\tau$ decay width.   By taking the  average upon this set, with a conservative treatment of the errors, we obtain   $\alpha_s(M_\tau^2)= 0.3195^{+ 0.0189}_{- 0.0138}$. 
\end{abstract}

\maketitle

\section{Introduction}\label{sec:intro}
The  conventional perturbation expansion of the QCD correlators in powers of the strong coupling $\alpha_{s}$ 
is problematic, because  the  function that is expanded, like for instance the Adler function ${\wh D}$, viewed as a function of the coupling, is known to be singular at the expansion point, $\alpha_s=0$ \cite{tHooft}. On the other hand, the powers 
$\alpha_s^{n}$ are holomorphic and, therefore, they can tell us nothing about the 
singularities of the expanded function, including their very existence. As a consequence, no finite-order 
perturbative approximant can share the singularity
with the expanded function at zero coupling. Singularities can emerge only from the infinite 
series as a whole, which, unfortunately, is not defined, since the perturbation series is divergent \cite{Muell1,Broa,Bene,Beneke}. 

A perturbation series would be more instructive if the individual 
finite-order approximants could retain at least some information about the 
known singularities of the QCD correlators. Such approximants would, in
every order of approximation, bear some information about the singularities of 
the expanded function and, moreover, would tell us more about the function
from the numerical point of view. 

An  approach proposed by us some time ago \cite{CaFi1}  consists in replacing 
the conventional set of powers of $\alpha_s$ (occurring in the standard 
perturbation expansion) by a new set of functions,  in which the
available information about the singularities 
of the expanded function is built in.  
 In order to define such a new perturbation series, two  methods can be used: 
(A) extension of the region of convergence by conformally mapping the region of holomorphy of the expanded function  onto a disk \cite{CiFi}, and (B) singularity softening, discussed for instance in \cite{SoSu, CaFi1}. When combined in a suitable way, they are mutually amplifying their effect. 
	
The method of conformal mapping was introduced and applied in particle physics in \cite{CiFi}, with an intent of extending the convergence region beyond the circle of convergence of an expansion and increasing the convergence rate at points lying inside the circle. In the context of perturbative QCD, the  properties of the new expansions based on this technique were investigated in \cite{CaFi1, CaFi2,CaFi3}, while in \cite{CaFi2009}  the method was applied for a determination of $\alpha_s$ from hadronic  $\tau$  decays. 

 The conformal mapping method  \cite{CiFi} is not applicable to the 
(formal) perturbative series of ${\wh D}$ in powers of  $\alpha_{s}$, because ${\wh D}$  is singular 
at the point of expansion\footnote{In the so-called "order-dependent" conformal mappings, which were defined also in the coupling plane \cite{ZJ, ZJ1}, the singularity is  shifted away from the origin by a certain amount at each finite-order, and tends to the origin only when an infinite number of terms are considered.}.  On the other hand, the method can be applied \cite{CaFi1}, rather than to   ${\wh D}$,  to its Borel transform $B(u)$ (the precise definition of this function will be given in Sec. \ref{sec:adler}). Being holomorphic in a region containing the origin  $u = 0$  of the Borel complex plane, $B(u)$  can be expanded in powers of  the Borel variable as
\be\label{eq:B} B(u)=\sum_{n\ge 0} b_n u^n. \ee
This series is convergent inside the circle centered at the origin  $u = 0$  and reaching the nearest singularity of  $B(u)$.  It often happens in practice that the disk of convergence of (\ref{eq:B}) is considerably smaller than ${\bf\cal B}$, the holomorphy domain of  $B(u)$ (we assume throughout the paper that ${\bf\cal B}$ is simply-connected). It is the method of conformal mapping (A) that can significantly extend the region of convergence of (\ref{eq:B}), by replacing the series  in powers of  $u$  with a series in powers of a new variable, $\wt w(u)$,
\be\label{eq:Bw} B(u)=\sum_{n\ge 0} c_n (\wt w(u))^n. \ee
Here  $\wt w(u)$  denotes the function that conformally maps the holomorphy region  ${\bf\cal B}$ in the  $u$-plane onto the unit disk $|w|<1$ in the  $w\equiv \wt w(u)$-plane centered at the origin (the explicit form of $\wt w(u)$ will be given in Sec. \ref{sec:confmap}).  The expansion coefficients  $c_n$  are determined by the  $b_n$ coefficients and by  the conformal mapping function  $\wt w(u)$.  The expansion  (\ref{eq:Bw}), unlike  (\ref{eq:B}), is convergent in the whole region of holomorphy  ${\bf\cal B}$. Moreover, as proved in Ref. \cite{CiFi} and will be discussed also in  Sec. \ref{sec:confmap} of the present paper, it   provides the fastest large-order convergence rate.

In the present work we focus on the procedure of  singularity softening, which exploits the known nature of the leading singularities of the correlators in the Borel plane, by compensating them with suitable factors. As discussed in \cite{CaFi1, CaFi2009, CaFi_RJP}, this procedure is not unique, in contrast with the definition of the optimal mapping \cite{CiFi}, which is unambiguous. 

 To illustrate the method, we consider  the Adler function in massless QCD.  In the next section we briefly review the properties of this function in perturbative QCD.  In Sec. \ref{sec:confmap} we discuss the method of conformal mapping  for improving the convergence rate of  power series \cite{CiFi}.   Although the method  has been  adopted and applied by many authors, its mathematical foundation is not very  widely known.  We therefore formulate two lemmas  which allow the definition of the so-called "optimal" conformal mapping in this context. The proof of the lemmas is presented in Appendix.

In Sec. \ref{sec:singsoft}, we discuss various possibilities of exploiting the known nature of the leading singularities of the Borel transform, and in Sec. \ref{sec:expfct}, we define a class of new expansion functions that implement the two ingredients, singularity softening and expansion in a new variable. Both the contour-improved and the fixed-order versions of the expansions are presented. In Sec. \ref{sec:models} we illustrate the  convergence properties of the new expansion functions using some mathematical models for the Adler function. 

 The determination of the strong coupling  $\alpha_s$ is one of the most important tests of QCD. As discussed in \cite{Bethke}, the recent determinations at various scales are in an impressive agreement among each other. The hadronic decays of the $\tau$ lepton provide one of the most precise determination, which is particularly interesting  as it concerns a relatively low scale, the mass of the $\tau$. 
The recent calculation of the Adler function to four loops \cite{BCK08}, the same order to which the $\beta$-function is known \cite{LaRi,Czakon}, renewed the interest in the determination of $\alpha_s(M_\tau^2)$ \cite{Davier2008}-\cite{Beneke_Muenchen}.  In  Sec.\ref{sec:alphas} we present an updated calculation of the strong coupling using the new CI expansion functions defined in this work.  Finally, Sec. \ref{sec:disc} contains a summary of the work and our conclusions.

\vskip-0.7cm

\section{Adler function}\label{sec:adler}
 We consider the Adler function  \cite{Adler}, {\em i.e.} the logarithmic derivative of the correlation function of two hadronic currents, which is expressed in  massless perturbative QCD as
\be \label{eq:Dpert}
\wh D(s) = \sum\limits_{n\ge 1} [K_{n}+\kappa_n(-s/\mu^2)]\,  (a_s(\mu^2))^n,
\ee
where  $s=q^2$ is the momentum variable and $a_s(\mu^2)\equiv \alpha_s(\mu^2)/\pi$ is the strong coupling at the renormalization scale $\mu^2$.  From studies of classes of Feynman diagrams it is known \cite{Broa,Bene,Beneke} that the series in the right hand side of (\ref{eq:Dpert}) is divergent. This series is often assumed \cite{Beneke} to be an asymptotic expansion in the limit $a_s\to 0$. Then the equality sign in (\ref{eq:Dpert}) is interpreted as $\sim$, the sign used for asymptotic  expansions. 

 The first coefficients  $K_n$ calculated in the $\overline{\rm MS}$ scheme are  \cite{BCK08} 
\be\label{eq:Kn}
K_1=1,~K_2= 1.63982,~K_3=6.37101,~K_4=49.0757. 
\ee
 Several estimates of the next coefficient $K_5$ are  available: the value $K_5=378$, obtained from the assumption of a geometrical growth, was adopted in \cite{Davier2008}, while the "Fastest Apparent Convergence" (FAC) principle \cite{Grun} predicts $K_5=275$ \cite{BCK08, Pich2010}. A slightly different value, $K_5=283$, was adopted in \cite{BeJa}.  

The coefficients $\kappa_n(-s/\mu^2)$ depend on the renormalization-group (RG) $\beta$-function, which is calculated at present to four loops \cite{LaRi, Czakon}. The first coefficients  $\beta_j$ in the  $\overline{\rm MS}$ scheme for $n_f=3$ are 
\be\label{eq:betaj}
\beta_0=9/4,~~\beta_1=4,~~\beta_2=10.0599,~~\beta_3=47.228.
\ee 
An additional term is  sometimes added to the perturbative expansion of the $\beta$-function, assuming a geometrical growth, $\beta_4 =\pm \beta_3^2/\beta_2$ \cite{Davier2008,Pich_Manchester}.

The Adler function plays a crucial role in the determination of  $\alpha_s(M_\tau^2)$ from hadronic $\tau$ decays.  The method is discussed in the seminal paper \cite{BrNaPi} and is reviewed in several recent articles \cite{Davier2008, BeJa, Pich_Manchester, Pich_Muenchen}. For completeness we give below a few details. 

The inclusive character of the total $\tau$ hadronic width makes possible  an accurate
calculation of the ratio  $R_\tau \,\equiv\,\Gamma[\tau^- \to \nu_\tau {\rm hadrons} \,  ]/
\Gamma[\tau^- \to \nu_\tau e^- \overline \nu_e ]$.  
Of interest is the Cabbibo allowed component which proceeds either through a 
vector or an axial vector current, since in this case the power corrections are
especially suppressed.  It can be expressed in the
form 
\begin{equation}
\label{eq:RtauVA}
R_{\tau,V+A} \,=\,N_c\,S_{\rm EW}\,|V_{ud}|^2\,\biggl[\,
1 + \delta^{(0)} + \delta_{\rm EW}' + \delta_{\rm PC} \,\biggr] \,,
\end{equation}
where $N_c=3$ is the number of colors,  $S_{\rm EW}$ and
$\delta_{\rm EW}'$  are electroweak corrections, $\delta_{\rm PC}$ denotes nonperturbative power corrections which arise in the framework
of the operator product expansion (OPE), and  $\delta^{(0)}$ is the genuine perturbative QCD correction.  Unitarity implies that  this quantity can be written as an integral over the spectral function of the polarization function along the timelike axis. As shown in \cite{BrNaPi}, the analytic properties of the polarization function and Cauchy theorem allow one to write  equivalently  $\delta^{(0)}$ as the contour integral 
\be\label{eq:delta0}
\delta^{(0)} =  \frac{1}{2\pi i}\, \oint\limits_{|s|=M_\tau^2}\, \frac{d s}{s}\, \left(1-\frac{s}{M_\tau^2}\right)^3\,\left(1+\frac{s}{M_\tau^2}\right) \wh D(s).\ee
As discussed in \cite{BrNaPi}, perturbative QCD is valid outside the timelike axis, so the Adler function can be calculated along the complex contour using the expansion (\ref{eq:Dpert}).

More generally, of interest are the moments of the spectral function, defined for arbitrary $s_0$ either as \cite{Neubert}
\be \label{eq:Mk}
M_k= \frac{1}{2\pi i}\, \oint\limits_{|s|=s_0}\, \frac{d s}{s}\, \left( 1-\frac{s^{k+1}}{s^{k+1}_0}\right) \, \wh D(s),~~~k\ge 0,
\ee
 or as \cite{MaYa}
\be \label{eq:barMk}
\bar{M}_k= \frac{1}{2\pi i}\, \oint\limits_{|s|=s_0}\, \frac{d s}{s}\, \left(1-\frac{s}{s_0}\right)^k \, \wh D(s), ~~k\ge 1.
\ee

The main ambiguity in the evaluation of these contour integrals is related to the renormalization scale. The choice  $\mu^2=-s$,  when (\ref{eq:Dpert}) reads
\be \label{eq:DpertCI}
\wh D(s) = \sum\limits_{n\ge 1} K_{n}\,  (a_s(-s))^n,
\ee
 leads to the so-called "contour-improved" (CI) expansion \cite{DiPi}, where the coupling is determined by solving the renormalization group equation exactly along the circle. The more conventional fixed-order (FO) expansion (\ref{eq:Dpert}) of $\wh D(s)$, when  $\mu^2=M_\tau^2$ (or more generally $\mu^2=s_0$), is obtained formally from (\ref{eq:DpertCI}) by  expanding the running coupling $a_s(-s)$ as
\be\label{eq:astaylor}
a_s(-s)=\sum\limits_{j\ge 1}\, \xi_j\, (a_s(M_\tau^2))^j.
\ee
The coefficients  $\xi_j$  depend on the parameters $\beta_k$, $k\le j$,  and the powers of $\ln (-s/M_\tau^2)$, which can acquire large imaginary parts for $s$ on the integration circle near the timelike axis. As discussed recently  \cite{BeJa, Pich_Manchester, Beneke_Muenchen}, the discrepancy between the results given by the CI and FO expansions is the main theoretical error in the extraction of  $\alpha_s(M_\tau^2)$. 

As already mentioned, the renormalized perturbation series (\ref{eq:Dpert}) or (\ref{eq:DpertCI}) are divergent, the coefficients displaying at large orders a factorial growth, $K_n\sim n!$. From independent arguments it is known that correlation functions like $\wh D$, regarded as functions of $\alpha_s$, are singular at $\alpha_s=0$ \cite{tHooft}. For QED, where these facts are well-known  \cite{Dyson}, the divergence of the series does not affect the phenomenological predictions since the coupling is very small. By contrast,  for a large coupling like $\alpha_s(M_\tau^2)$ in  QCD  the consequences are nontrivial.  

The information about the high-order behavior of the series (\ref{eq:Dpert}) is included in the singularities of the Borel transform $B(u)$, defined by the series (\ref{eq:B}), with the coefficients $b_n$  related to $K_n$ by
\be\label{eq:Kntobn}  b_n=\frac{K_{n+1}}{\beta_0^n \,n!}\,,\quad n\ge 0, \ee
where $\beta_0$ is the first coefficient of the $\beta$-function given in (\ref{eq:betaj}). According to present knowledge \cite{Beneke},  $B(u)$ has singularities on the real axis  for $u \le -1$ and $u\ge 2$, known as  ultraviolet (UV) and infrared (IR)  renormalons, respectively. In the present paper we assume there are no other singularities in the complex plane, so that the holomorphy domain ${\bf\cal B}$ is the $u$-plane cut along the real axis for $u\le -1$ and $u\ge 2$.

The expansions (\ref{eq:Dpert}) and (\ref{eq:DpertCI}) can be formally obtained from $B(u)$ by means of an integral of Borel-Laplace type. The recovery of the function $\wh D(s)$ is actually ambiguous: there are many integral representations admitting (\ref{eq:Dpert}) or (\ref{eq:DpertCI}) as  asymptotic expansions (for a recent discussion, see \cite{CaFiVr}). As shown in \cite{CaNe}, the definition based on the principal value prescription,
\be\label{eq:pv}
\wh D(s)=\frac{1}{\beta_0}\,{\rm PV} \,\int\limits_0^\infty  e^{-\frac{u}{\beta_0 a_s(-s)}} \, B(u)\, {\rm d} u,
\ee
yields a function $\wh D(s)$ satisfying to a large extent the general analyticity requirements in the $s$-plane, and we shall adopt this definition.

\section{ Accelerating convergence by conformal mappings}\label{sec:confmap}

Because of the first UV singularity at $u=-1$, the  expansion (\ref{eq:B}) 
converges only on the disk $|u| < 1$, although $B(u)$  is  holomorphic in a much
  larger region.
The domain of convergence and the 
convergence rate can be increased by expanding the function in powers of a 
different variable, defined by the conformal mapping of ${\bf\cal B}$ (the cut $u$-plane), or a part of it, onto a disk (without loss of generality the disk can be taken of radius equal to unity, and we shall adopt this convention).
It may intuitively seem that the larger the domain 
mapped onto the unit disk, 
the better the convergence properties of the series expansion in powers of the new variable. 
This is indeed true, and we shall give this hope a precise mathematical form.  The result, proved in Ref. \cite{CiFi}, important and interesting as it 
is for a number of applications, did not raise enough interest as it deserved, 
in spite of the many applications of the conformal mapping method during the 
decades. We shall therefore state below the main ideas of the proof, 
in order to make the present paper  self-contained. 
 The following two lemmas  show which is the variable that provides the best asymptotic rate of convergence.

\vspace{0.2cm}\noindent{\it Lemma 1:}~  Let ${\cal D}_1$  and  ${\cal D}_2$  be two  domains in the complex $u$-plane, with $ {\cal D}_2 \subset {\cal D}_1$, $ {\cal D}_2 \ne {\cal D}_1$, such that the conformal mappings  ${\cal D}_1\to {\cal K}_1$ and  ${\cal D}_2\to {\cal K}_2$ exist, where ${\cal K}_1$ and  ${\cal K}_2$ are unit disks. Consider the two conformal mappings
\bea\label{eq:z1z2}
z_1=\tilde z_1(u): {\cal D}_1\to {\cal K}_1=\{z_1:|z_1|<1\},\nonumber\\
 z_2=\tilde z_2(u): {\cal D}_2\to {\cal K}_2=\{z_2:|z_2|<1\}.
\eea
Let $Q$ be a point of ${\cal D}_2$,  $Q\in {\cal D}_2$, such that $ \tilde z_1(Q)=0$ and $\tilde z_2(Q)=0$. 
Then 
\be\label{eq:lema1} 
|\tilde z_1(u)| <  |\tilde z_2(u)|, \quad {\rm for~ all}~~ u\in {\cal D}_2, ~~u\ne Q. 
\ee

\vspace{0.2cm}\noindent
{\it Lemma 2:}~  Let ${\cal D}_1$  and  ${\cal D}_2$ be the domains defined in Lemma 1, $\tilde z_1(u)$ and  $\tilde z_2(u)$ the mappings (\ref{eq:z1z2}) and $B(u)$ a function holomorphic in ${\cal D}_1$. Define the expansions
\be\label{eq:z1}
 B(u)=\sum_{n=0}^{\infty} c_{n,1} (\tilde z_1(u))^n,
\ee
\be\label{eq:z2}
 B(u)=\sum_{n=0}^{\infty} c_{n,2} (\tilde z_2(u))^n, 
\ee
which are convergent for $z_1\equiv \tilde z_1(u) \in {\cal K}_1$ and $z_2\equiv \tilde z_2(u) \in {\cal K}_2$, respectively. 
Assume in addition that the limits $\lim_{n \to \infty} \root n \of{|c_{n,1}|}$ and $\lim_{n \to \infty} \root n \of{|c_{n,2}|}$ exist\footnote{The essence of this  is that the expansions (\ref{eq:z1}) and  (\ref{eq:z2})  have equal radii of convergence. This assumption is nontrivial, because the expanded (Adler) function might be of such a form that certain singularities of  $B(u)$  in  $z_1$  or  $z_2$  might disappear.} and are equal to one:
\be\label{eq:limsup}
\lim_{n \to \infty} \root n \of{|c_{n,1}|} =\lim_{n \to \infty} \root n \of{|c_{n,2}|}=1.
\ee
 Then a positive integer $N=N(u)$ exists such that the following inequality holds:
\be\label{eq:rate}
{\cal R}_n(u)=\left\vert\frac{c_{n,1} (\tilde z_1(u))^n}{c_{n,2} (\tilde z_2(u))^n}\right\vert <1,
\ee
for any $n$ integer, $n>N$, and $u\in {\cal D}_2$, $u\ne Q$.

Proofs of Lemma 1 and Lemma 2  are presented in Appendix.

\subsection{Optimal conformal mapping}
It is now easy to understand from Lemma 1 and Lemma 2 that the larger the domain mapped onto the disk, the larger the domain where the expanded function is represented by a convergent power series and, also, the faster the convergence rate at a given point. From the inequality (\ref{eq:rate}),  it follows that the best asymptotic convergence rate is obtained with the variable $\wt w(u)$  that maps ${\bf\cal B}$, the holomorphy domain of  $B(u)$, onto the unit disk $|w|<1$ in the plane $w\equiv \wt w(u)$.  In this case, the boundary of ${\bf\cal B}$  is mapped on  the boundary circle of the unit disk and the series (\ref{eq:Bw}) in powers of   $\wt w(u)$ is convergent everywhere in ${\bf\cal B}$. Moreover,  the asymptotic convergence rate of this series  is, at any point  $u\in {\bf\cal B}$, the fastest among all conformal mappings. This mapping is known as "optimal" conformal mapping for convergence acceleration \cite{CiFi,CaFi1, CaFi2009}.

Let us discuss an example to illustrate this result. Assume we decide to modify this mapping by adding a  region lying outside the holomorphy domain. In doing so, we wedge a region containing singularities inside the circle and unavoidably make the convergence radius smaller. As a consequence, the large-order convergence rate is worse. 

If, on the other hand, we omit to map a part of the holomorphy region inside the unit circle, the convergence rate becomes worse, as follows by a direct application of (\ref{eq:rate}). The reason is that we do not make full use of analyticity in this case, leaving aside a part of the holomorphy region. 
We conclude that by $\wt w(u)$, the optimal conformal mapping function, neither any singularity is mapped inside the circle, nor any part of the holomorphy region is left out of the circle.

As can be seen from the proof of Lemma 2, when $n$ is large enough, the inequality (\ref{eq:lnr}) reduces  to $\ln \rho(u)<0$, {\em i.e.} the coefficients $c_{n,j}$  play no role in the ratio ${\cal R}_n(u)$ of the convergence rates. On the other hand, when $n$ is  finite (and small), the term $g(n)/n$ in  (\ref{eq:lnr}) may be positive and greater than  $\ln \rho(u)$, making the ratio ${\cal R}_n(u)$ greater than one. This may happen, in particular, when both $|\tilde z_1(u)|$ and $|\tilde z_2(u)|$, as well as their ratio  $\rho$, are close to 1. Therefore, the expansions in powers of other conformal mappings can provide at finite orders a better approximation compared to the optimal expansion, a fact observed numerically  in some cases, especially for points near the boundary of the analyticity  domain. 

For the Adler function, assuming that there are no other singularities except for the cuts along the real axis for $u\le -1$ and $u\ge 2$, the optimal conformal mapping is \cite{CaFi1}
\be\label{eq:w}
\wt w(u)=\frac{\sqrt{1+u}-\sqrt{1-u/2}}{\sqrt{1+u}+\sqrt{1-u/2}},\ee
with the inverse 
\begin{equation}\label{eq:winv}
\wt u(w) = \frac{8 w}{3 - 2 w + 3 w^2}. 
\end{equation}

Using the optimal expansion (\ref{eq:Bw}) and the definition (\ref{eq:pv}), we were led in a natural way to the new perturbative expansion \cite{CaFi1}
\be\label{eq:DnewCI}
\wh D(s)=\sum\limits_{n\ge 0} c_n {\cal W}_n(s), 
\ee
\vspace{-0.1cm}
\be\label{eq:Wn}
{\cal W}_n(s)=\frac{1}{\beta_0}{\rm PV} \int\limits_0^\infty\! \,{\rm e}^{-\frac{u}{\beta_0 a_s(-s)}}  (\wt w(u))^n\,{\rm d}u.
\ee
By construction, the series (\ref{eq:DnewCI}), when reexpanded in powers of $\alpha_s$, reproduces the expansion (\ref{eq:DpertCI}) with the coefficients $K_n$   known from Feynman diagrams.  On the other hand,  the expansion functions ${\cal W}_n$ are singular at $\alpha_s=0$, resembling the expanded function $\wh D$ itself \cite{CaFi3}.  Moreover, as discussed in \cite{CaFi2}, under certain conditions, the expansion (\ref{eq:DnewCI}) converges in a domain of the $s$-plane.

\begin{figure*}[!ht]\includegraphics[width=4.2cm]{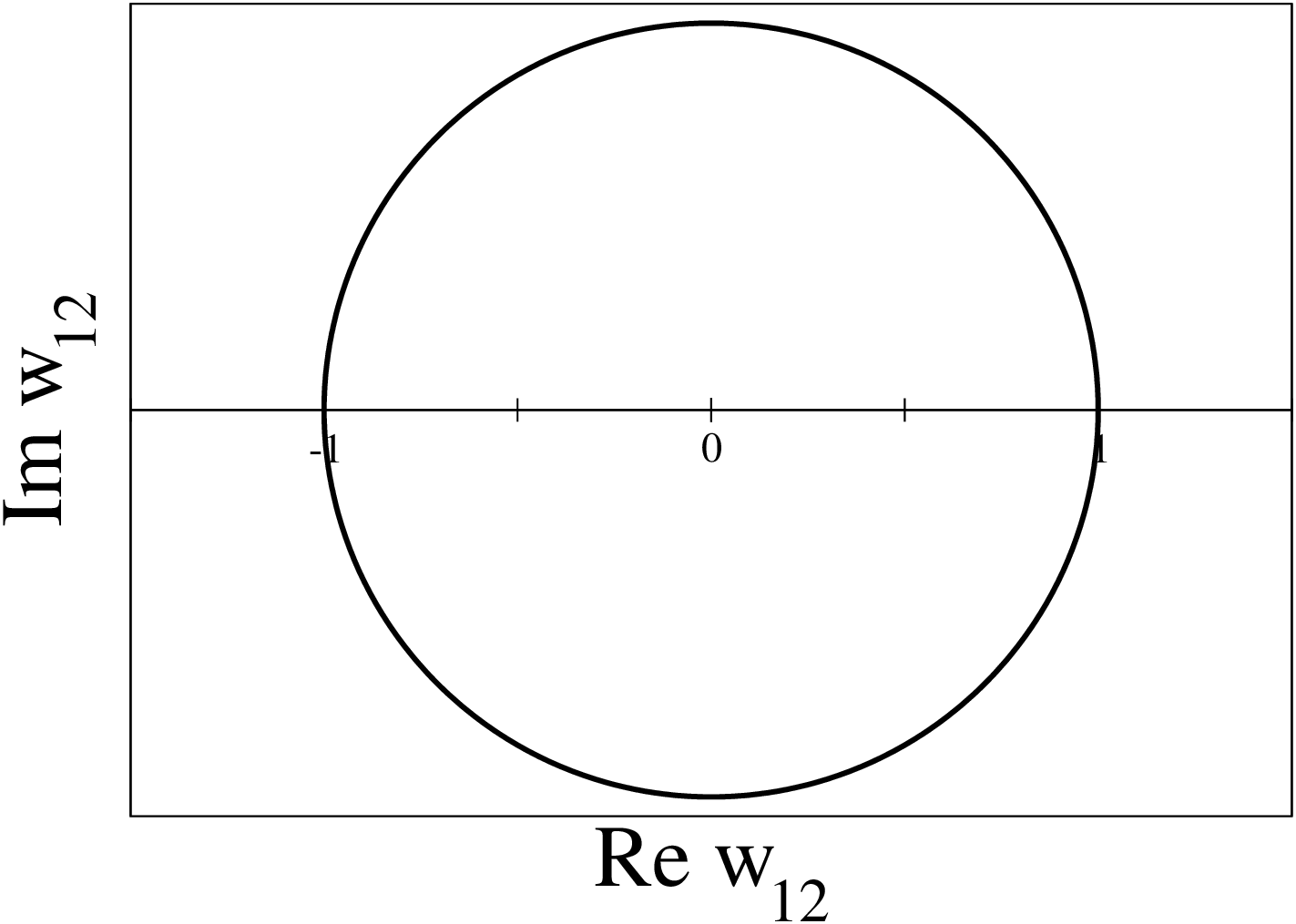}\hspace{0.3cm}
\includegraphics[width=4.2cm]{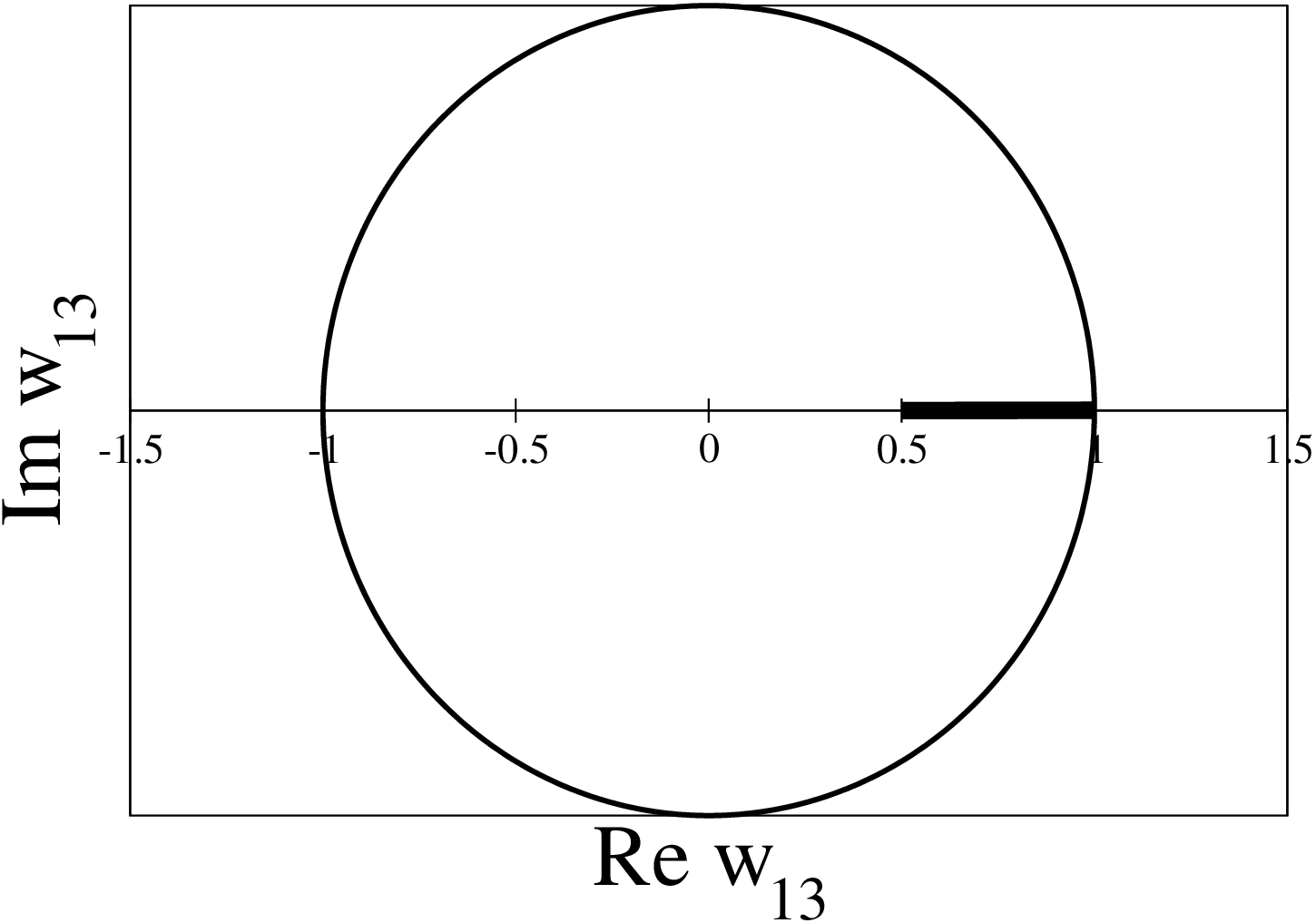}\hspace{0.3cm}\includegraphics[width=4.2cm]{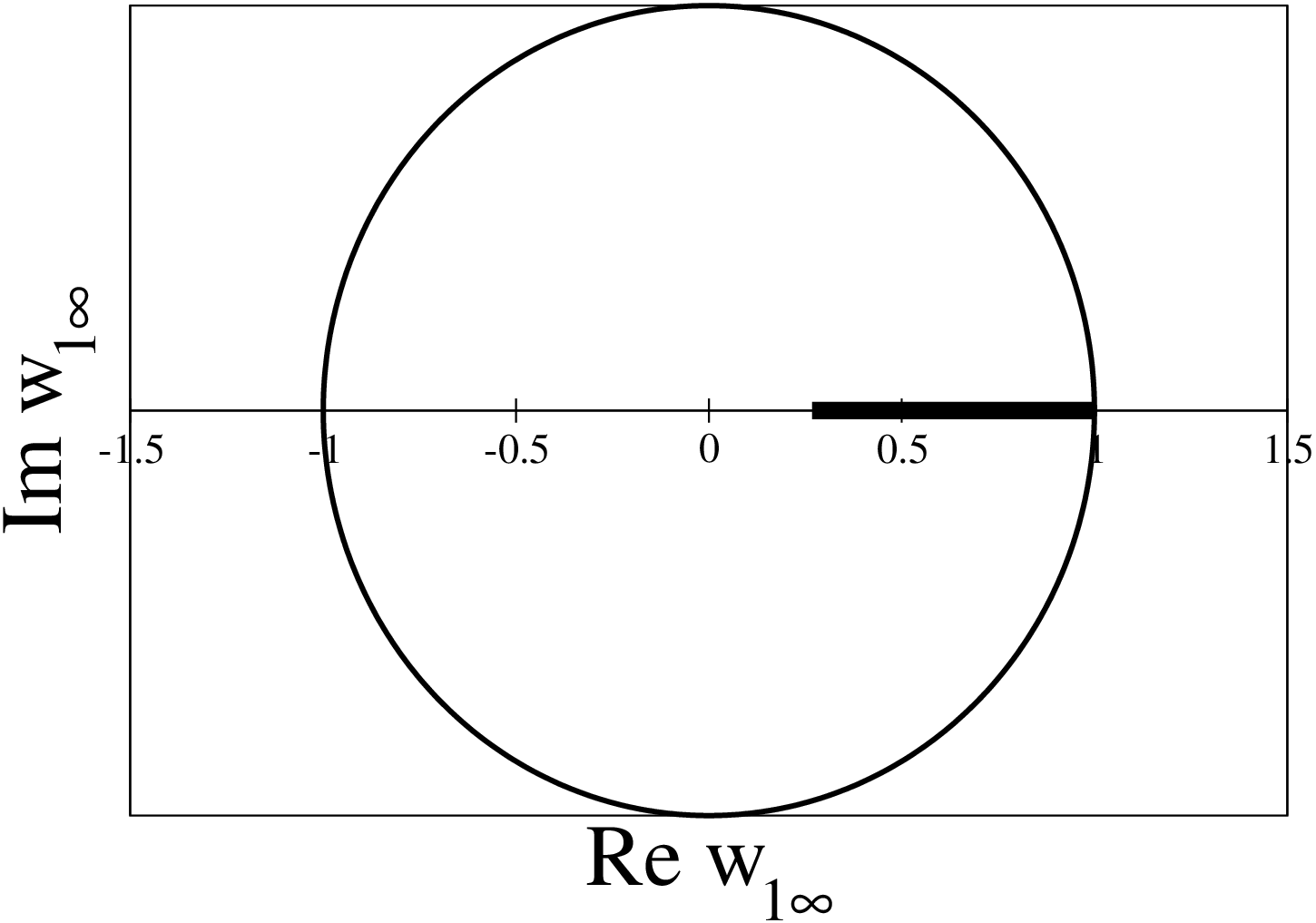}\hspace{0.3cm}\includegraphics[width=4.2cm]{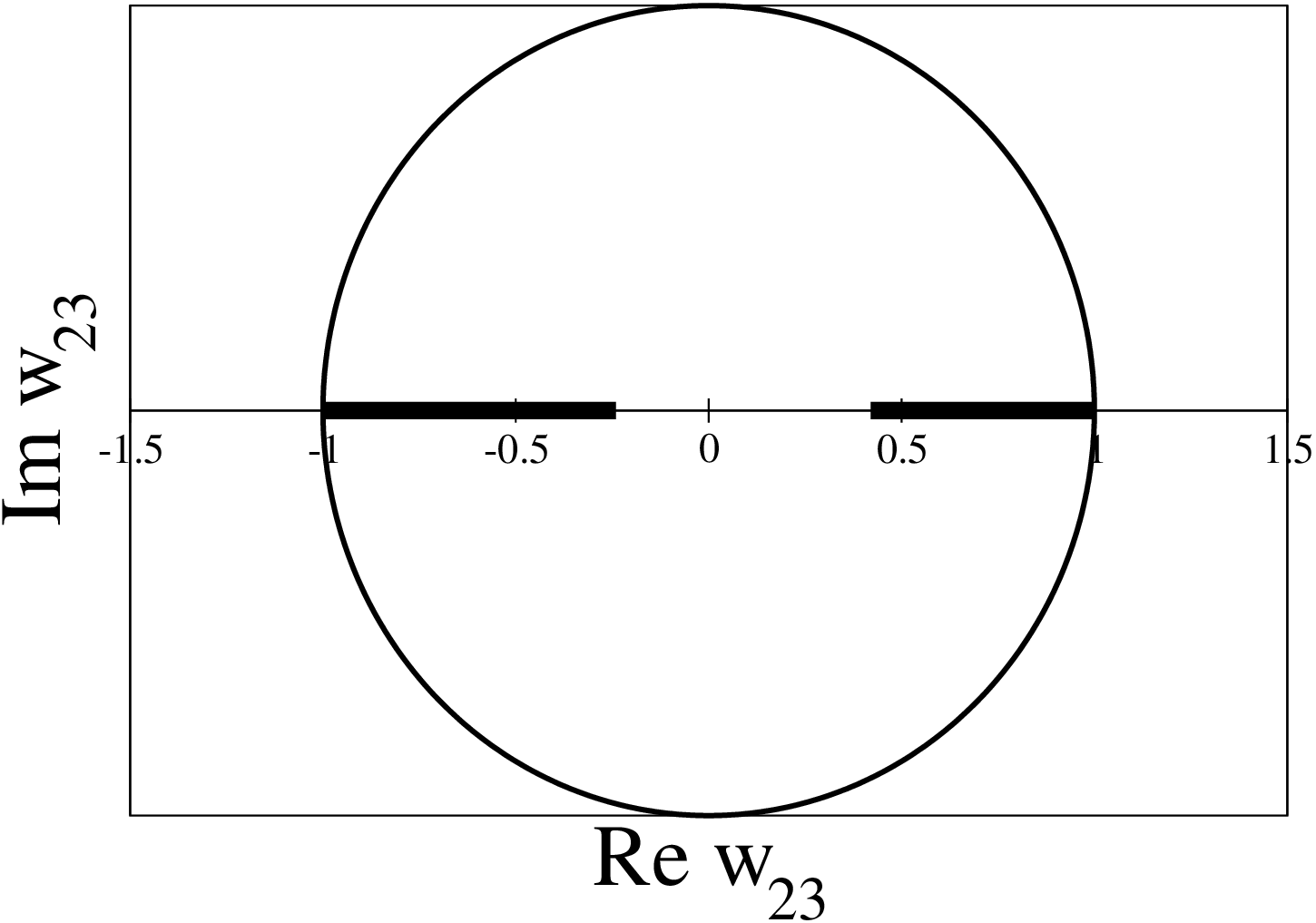}
\caption{\label{fig:wjk}   The unit disks $|w_{jk}|<1$ on which the conformal mapping defined in (\ref{eq:wjk}) maps the cut $u$-plane, for several values of $j$ and $k$. In the last three figures, the thick lines indicate the residual cuts along the segments $(\wt w_{13}(2), 1)$, $(\wt w_{1\infty}(2), 1)$ and $(-1, \wt w_{23}(-1)), ~ (\wt w_{23}(2), 1) $, which correspond, in the $u$-plane, to the segments $(2,3)$, $(2, \infty)$, and $(-2,-1)$, $(2,3)$, respectively.  }
\end{figure*}

\section{Singularity softening}\label{sec:singsoft}
In the particular case of the Adler function in massless QCD, the 
nature of the leading singularities in the Borel plane is known: 
near the first branch points, $u=-1$ and $u=2$,  $B(u)$ behaves like
\begin{equation}\label{eq:gammapowers}
B(u) \sim \frac{r_1}{(1+u)^{\gamma_{1}}} \quad{\rm and} \quad B(u)  \sim \frac{r_2}{(1-u/2)^{\gamma_{2}}}, 
\end{equation}
respectively. The residues $r_1$ and $r_2$ are not known, but the exponents
$\gamma_1$ and $\gamma_2$,  calculated using renormalization-group
invariance, have known values  \cite{Muell,BBK,BeJa}
\begin{equation}\label{eq:gamma12}
\gamma_1 = 1.21,    \quad\quad   \gamma_2 = 2.58 \,. 
\end{equation}
The expansion (\ref{eq:Bw}) takes into account only the position 
of the renormalons in the Borel plane. If a sufficient number of expansion 
coefficients were known, (\ref{eq:Bw}) would be expected to describe also the 
character, strength, etc., of the singularities as well. Since, however, only a few
 perturbative coefficients are at present explicitly available, 
one cannot expect that the expansion of the type (\ref{eq:Bw}) might be 
able to give a satisfactory approximation of $B(u)$. It is better than 
(\ref{eq:B}), which has no singularities in any finite-order approximation, while
every finite-order approximant to (\ref{eq:B}) has the same location of cuts as
the expanded function. But it can hardly be expected that the first four or five
perturbative coefficients would be able to represent $B(u)$ with a satisfactory
accuracy. 

An explicit account for the leading singularities (\ref{eq:gammapowers}) would
therefore be helpful to further improve the convergence. This can be done by multiplying
$B(u)$ with suitable factors that vanish at $u=-1$ and $u=2$ and compensate the dominant singularities.
The subsequent expansion of the product in powers  of a conformal mapping variable is expected to 
converge better. This procedure is known as  "singularity softening" \cite{SoSu, CaFi1}.  

In contrast with the optimal conformal mapping, singularity softening is not unique  \cite{CaFi2009, CaFi_RJP, CaFi_Manchester}. 
The singularities are 
present in $B(u)$, but we do not know their actual form, except for the behavior (\ref{eq:gammapowers}) near 
the corresponding branch-points. A possibility is to multiply $B(u)$ by
simple factors like $(1+u)^{\gamma_1} (1-u/2)^{\gamma_2}$   \cite{SoSu,CaFi1}.
In \cite{CaFi2009} the alternative softening factors $(1+w)^{2 \gamma_1} (1-w)^{2 \gamma_2}$ were adopted, 
where $w=\wt w(u)$ is the optimal mapping (\ref{eq:w}). The product of $B(u)$ with these factors was afterwards expanded in powers of the same variable $w$. Other possibilities will be investigated in the next section.

\section{New expansion functions}\label{sec:expfct}
The product of $B(u)$ with softening factors is expected to contain milder singularities,  which vanish
instead of exploding at  $u=-1$ and $u=2$  (in very peculiar cases the singularities may disappear altogether, but this situation is very unlikely).  The effect of a mild singularity in a function is not visible at low orders  in its series expansions, and is  expected to appear only at large orders.  Therefore, we can ignore their effects, expanding  the product 
in powers of variables that account only for the next branch-points of $B(u)$. In the case of the Adler function, these singularities are placed at $u=3,\,4$, etc., on the positive axis, and at  $u=-2,\,-3$, etc., on the negative axis.

In general, we consider the functions
\be\label{eq:wjk}
\wt w_{jk}(u)=\frac{\sqrt{1+u/j}-\sqrt{1-u/k}}{\sqrt{1+u/j}+\sqrt{1-u/k}},\ee
 which map the $u$-plane cut along $ u\le -j$ and $u\ge k$ to the disk $|w_{jk}|<1$ in the plane $w_{jk}\equiv \wt w_{jk}(u)$.
For $j= 1$, $k= 2$, we recover the optimal mapping (\ref{eq:w}).  In the following, we shall consider also the variables $w_{12}$, $w_{13}$, $w_{1\infty}$ and  $w_{23}$, for which the corresponding unit disks $|w_{jk}|<1$ are shown  in Fig. \ref{fig:wjk}.  We mention that the mapping $w_{1\infty}$, suggested in \cite{Muell1}, was discussed in a similar context in \cite{Alta}, and  $w_{13}$ was investigated in \cite{CvLe}. According to the discussion in the previous section, the last three mappings "push" inside the unit circle a part of the $u$-plane containing some singularities (indicated in Fig.  \ref{fig:wjk}). As a consequence, the expansions based on these variables will converge in a smaller domain and their convergence rates  will be, in principle, worse than that of the optimal mapping $w_{12}$.

According to the above discussion, we shall expand in powers of  $w_{jk}$ the product of $B(u)$ with suitable softening factors. Specifically, we consider the expansions
\be \label{eq:prod} S_{jk}(u) B(u) = \sum_{n\ge 0} c_n^{jk}  (\wt w_{jk}(u))^n,
\ee
where  $S_{jk}(u)$ must "soften" in principle all the singularities of $B(u)$ at  $-j\leq u<0 $ and $0<u \leq k$.
 Numerically, it appears to be convenient to choose the factor $S_{jk}$ as a simple expression with a 
rapidly converging expansion in powers of $w_{jk}$, thus ensuring  a good convergence of the product (\ref{eq:prod}).

A systematic application of this idea to the singularities of $B(u)$ requires the knowledge of the nature of the branch-points, which at present is limited to the leading singularities at $u=-1$ and $u=2$.  
 Therefore, we shall limit ourselves to compensating factors that vanish at these points, and take  $S_{jk}$ of the form:
 \be\label{eq:Sjk}
 S_{jk}(u)=\left(\!1-\frac{\wt w_{jk}(u)}{\wt w_{jk}(-1)}\!\right)^{\!\!\gamma^{(j)}_1} \!\!\left(\!1-\frac{\wt w_{jk}(u)}{\wt w_{jk}(2)}\!\right)^{\!\!\gamma^{(k)}_2}. \ee
The exponents $\gamma_1^{(j)}=\gamma_1 (1+\delta_{j1})$ and 
 $\gamma_2^{(k)}= \gamma_2 (1+\delta_{k2})$, where $\delta_{ij}$ is Kronecker's function, are taken such as to reproduce the nature of the first branch-points of $B(u)$, given in (\ref{eq:gammapowers}). In particular, for the optimal case $j=1$, $k=2$ we recover from (\ref{eq:Sjk}) the factor $(1+w)^{2 \gamma_1}(1-w)^{2 \gamma_2}$ used in \cite{CaFi2009}, with $w=\wt w(u)$ defined in (\ref{eq:w}).

Strictly speaking, for a fixed pair ($j, k$) the expansion (\ref{eq:prod})  converges only on the disk $|w_{jk}|< \min [|\wt w_{jk}(-1)|,\, |\wt w_{jk}(2)|]$. For the optimal choice $j=1, k=2$, the expansion converges in the whole unit disk $|w_{12}|<1$, {\em i.e.} in the whole $u$-plane except for the cuts along the real axes for $u\ge 2$ and $u\le -1$ \cite{CaFi1}.  For other mappings, the convergence disk is limited by the beginning of the cuts shown in Fig. \ref{fig:wjk}. In particular,  if $j=1$ and $k>2$ the expansions (\ref{eq:prod})  diverge for real $u$ greater than 2,  while for the conformal mappings with $j>1$,  the expansions start to diverge for $u$ greater than one, due to the singularity at $u=-1$ pushed inside the circle (as in the last case shown in Fig. \ref{fig:wjk}).  However, the product $S_{jk}(u) B(u)$ has only mild singularities.
Moreover, the expansion (\ref{eq:prod})  enters the Laplace-Borel integral (\ref{eq:pv}) where,  for values of  $a_s$ in the domain of interest, the contribution of high values of $u$ is suppressed. In particular, if $a_s$ is not very large, the region $u>2$ brings a small contribution to the integral, so signs of divergence in the case of the variables  $w_{13}$ and  $w_{1\infty}$ are expected to occur only at very large orders $N$. On the other hand, for the variable $w_{23}$, it is natural to expect signs of divergence  at lower values of $N$, since the series (\ref{eq:prod}) does not converge for $u>1$ . 

By combining the expansion (\ref{eq:prod}) with the definition (\ref{eq:pv}), we are led to the class of expansions
\be\label{eq:Djk} \wh D (s) = \sum\limits_{n\ge 0} c_n^{jk} \, {\cal W}^{jk}_n(s),\ee
\be\label{eq:Wnjk} {\cal W}^{jk}_n(s)=\frac{1}{\beta_0} {\rm PV}\int\limits_0^\infty\!{\rm e}^{-\frac{u}{\beta_0 a_s(-s)}} \,\frac{(\wt w_{jk}(u))^n}{S_{jk}(u)}\, {\rm d} u.\ee
By inserting into (\ref{eq:Wnjk}) the coupling $a_s(-s)$ calculated by solving the renormalization-group equation for $s$ along the circle defined in the integral (\ref{eq:delta0}), we obtain  the countour-improved
(CI) version of the new expansions. 

The new fixed-order (FO)  expansions can be obtained in a straightforward way \cite{CaFi2009}, using as starting point (\ref{eq:Dpert}). They have the generic form 
\be\label{eq:DFOjk} \wh D(s) = \sum\limits_{n\ge 0} \tilde c_n^{jk}(s) \, \tilde{\cal W}^{jk}_n,\ee
\be\label{eq:WnFOjk} \tilde{\cal W}^{jk}_n=\frac{1}{\beta_0} {\rm PV}\int\limits_0^\infty\!{\rm e}^{-\frac{u}{\beta_0 a_s(M_\tau^2)}} \,  \frac{(\wt w_{jk}(u))^n}{S_{jk}(u)}\, {\rm d} u.\ee

The  expansion functions ${\cal W}^{12}_n$ coincide with the optimal functions investigated in detail in \cite{CaFi2009}. In the following, we shall  also consider the expansion in terms of the functions ${\cal W}^{13}_n$, ${\cal W}^{1\infty}_n$ and ${\cal W}^{23}_n$ (and their corresponding FO versions).  We emphasize that these  expansions  contain  different softening factors, which coincide only for $u$ near the corresponding singularities, when they reproduce the known behavior (\ref{eq:gammapowers}). The treatment of the residual singularities after softening is also different: the expansion in powers of the optimal mapping uses the position of the first singularities, which in general do not disappear completely after the multiplication with the compensating factors.  The other expansions  exploit the  fact that a mild singularity can be neglected at low  perturbative orders, and use also  some information about the position of higher renormalons. So, the representations 
(\ref{eq:Djk})-(\ref{eq:WnFOjk}) for different $j$ and $k$ can be considered  independent  perturbative expansions of the Adler function.

\begin{figure*}[!ht]\begin{center}\includegraphics[width=6.cm]{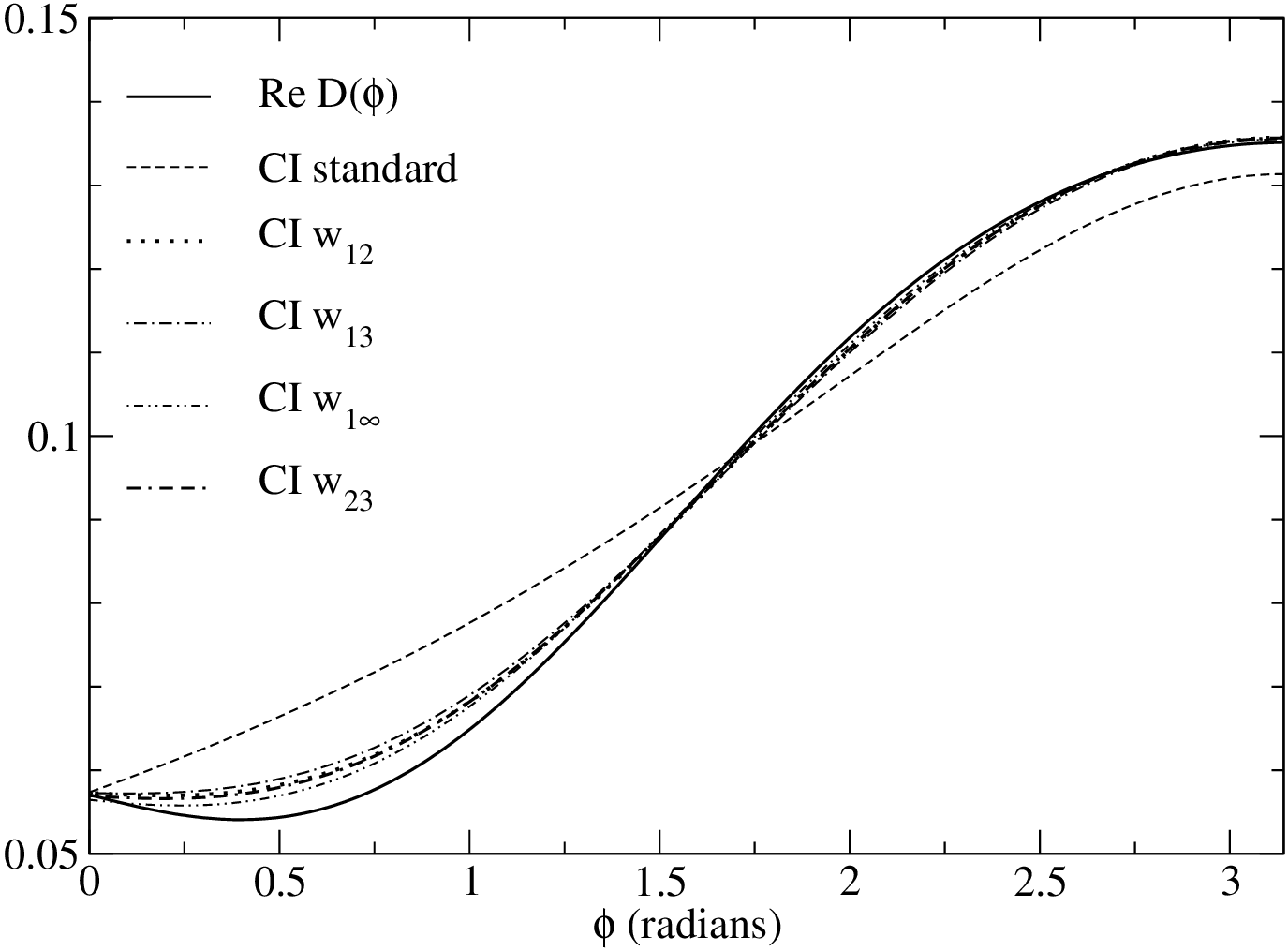}\hspace{1cm}
\includegraphics[width=6.cm]{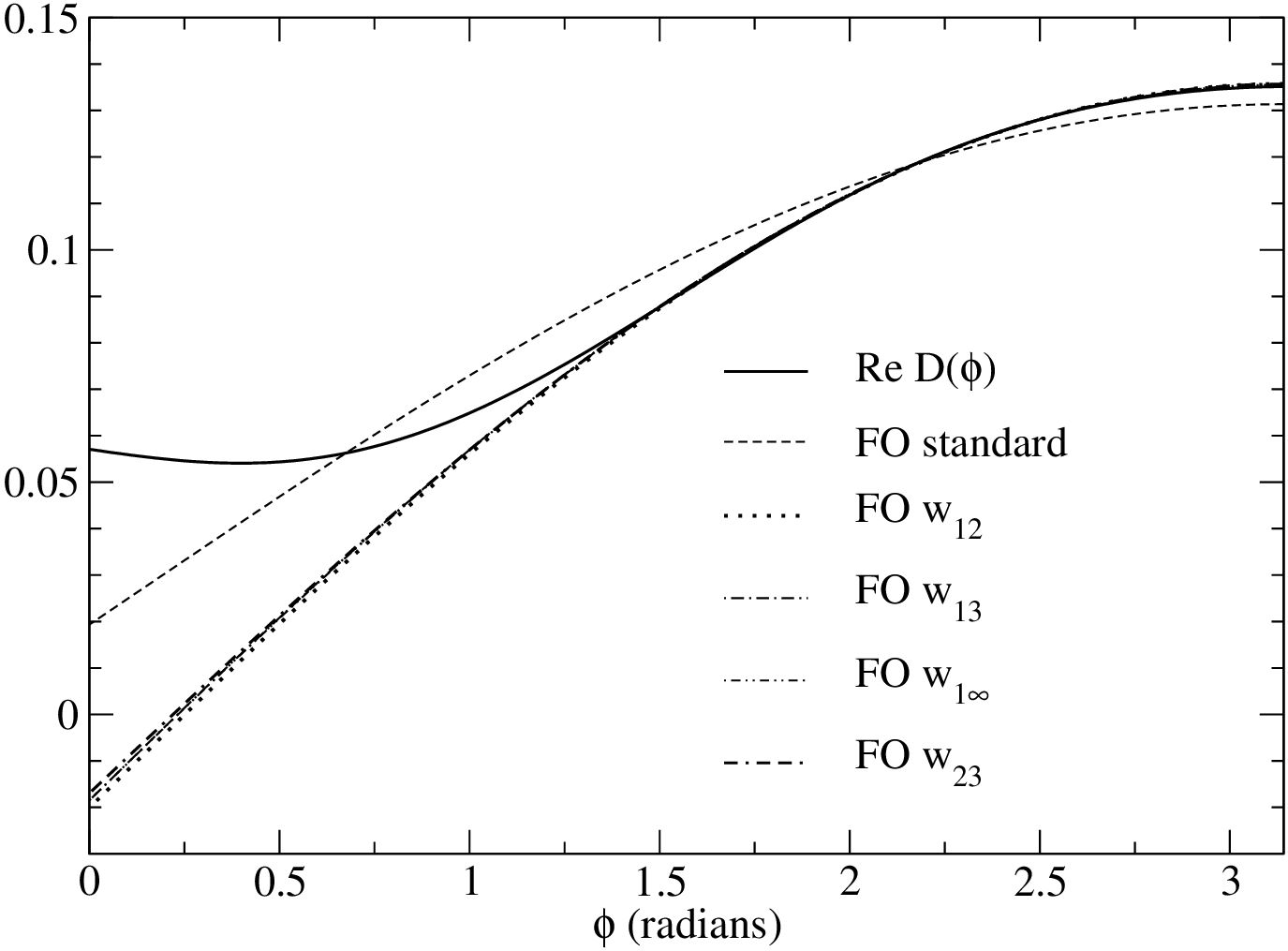}
\caption{\label{fig:DR5} Real part of the Adler function of the model \cite{BeJa} defined in (\ref{eq:BBJ})-(\ref{eq:dBJ}), calculated along the circle $s=M_\tau^2 \exp(i\phi)$ for $\alpha_s(M_\tau^2)= 0.3156$, using the perturbative expansions with $N=5$ terms. Left panel: CI expansions. Right panel: FO expansions. The exact function is represented by the solid line.}\end{center}\vspace{0.5cm}
\end{figure*}
\begin{figure*}[!ht]\begin{center}\includegraphics[width=6.cm]{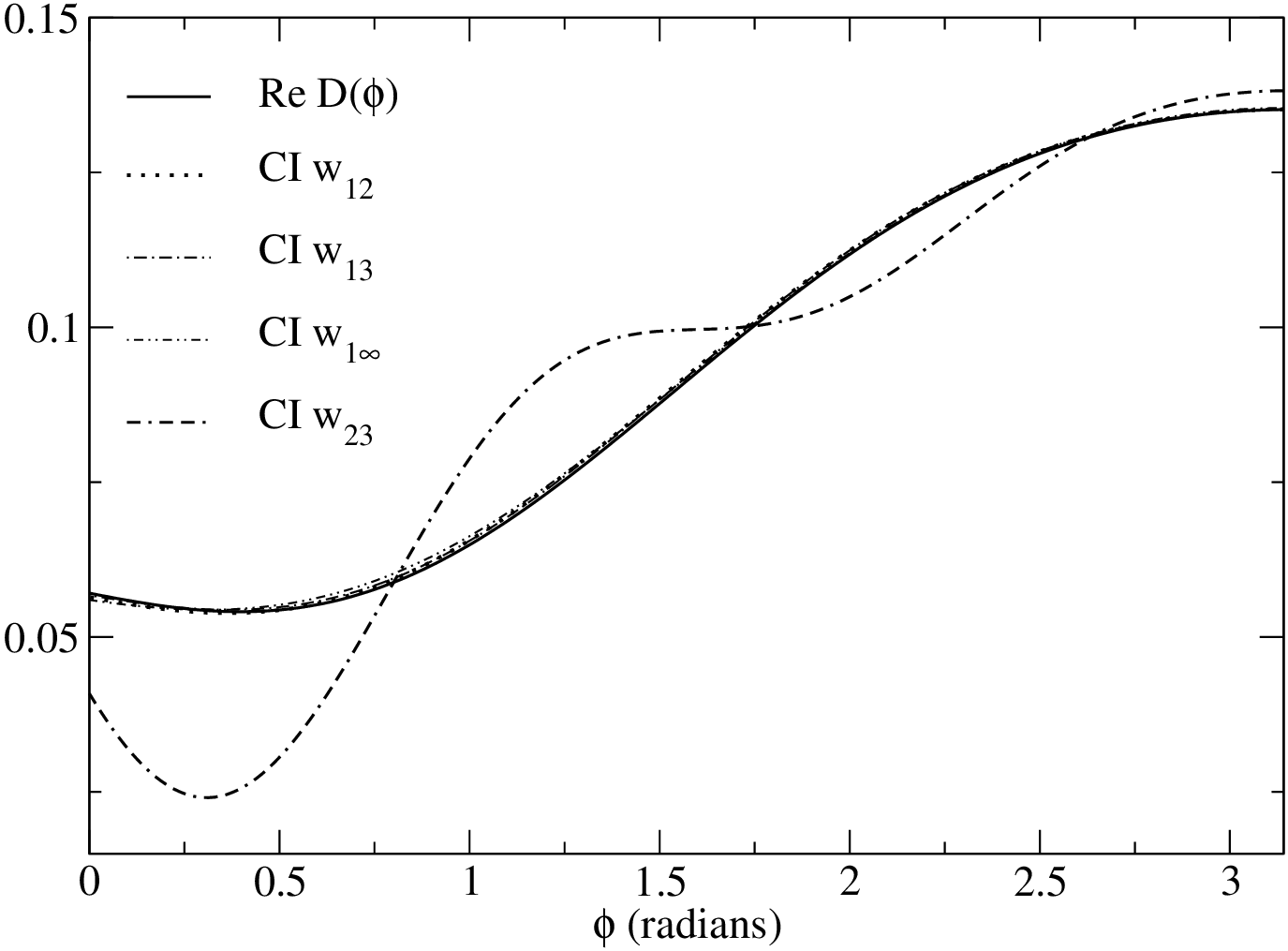}\hspace{1cm}\includegraphics[width=6.cm]{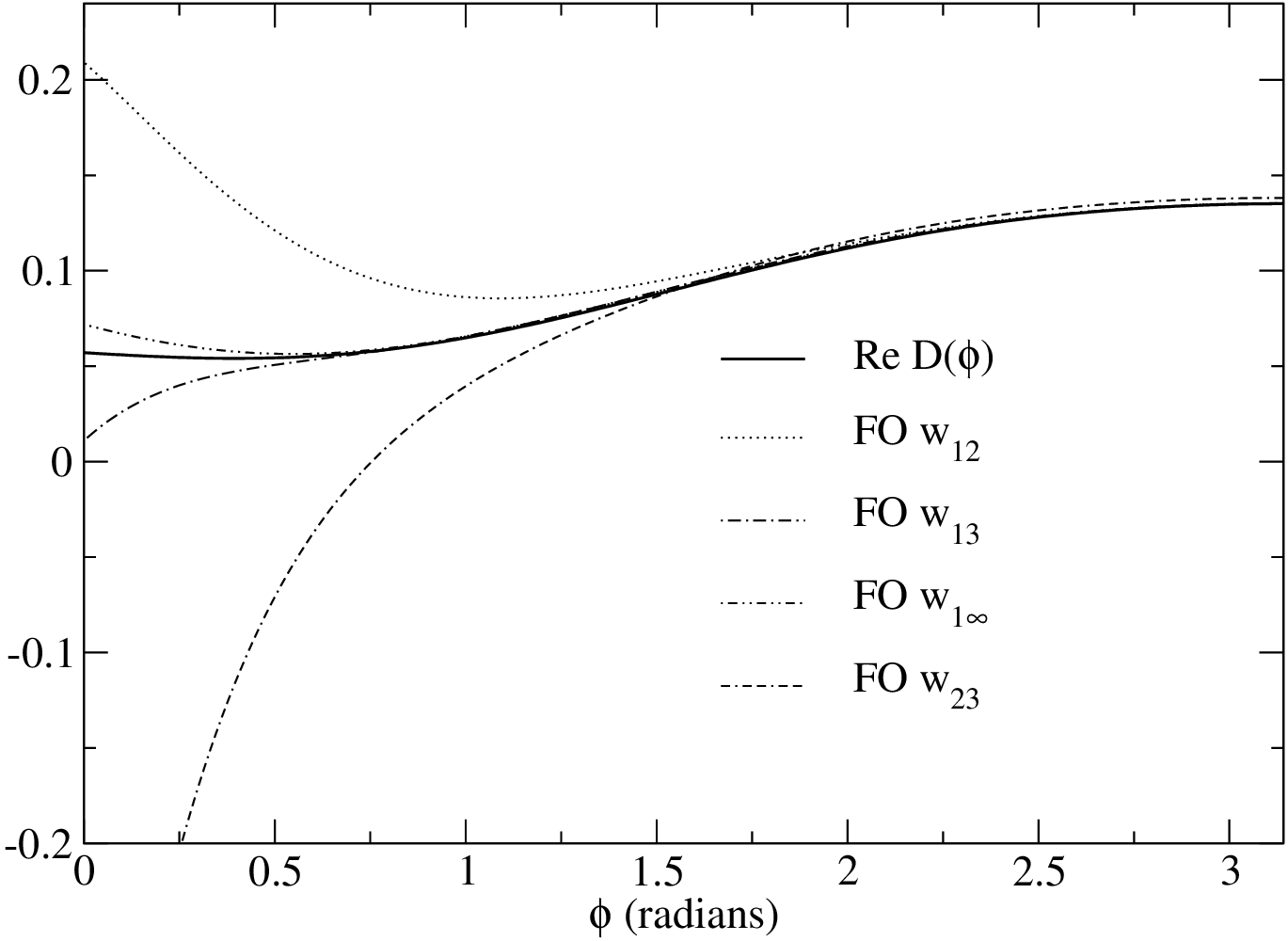}
\caption{\label{fig:DR18} As in Fig. \ref{fig:DR5} for $N=18$. The standard CI and FO expansions exhibit big oscillations and are not shown.}\end{center}\vspace{0.5cm}
\end{figure*}

\begin{figure*}[!ht]\begin{center}\includegraphics[width=6.cm]{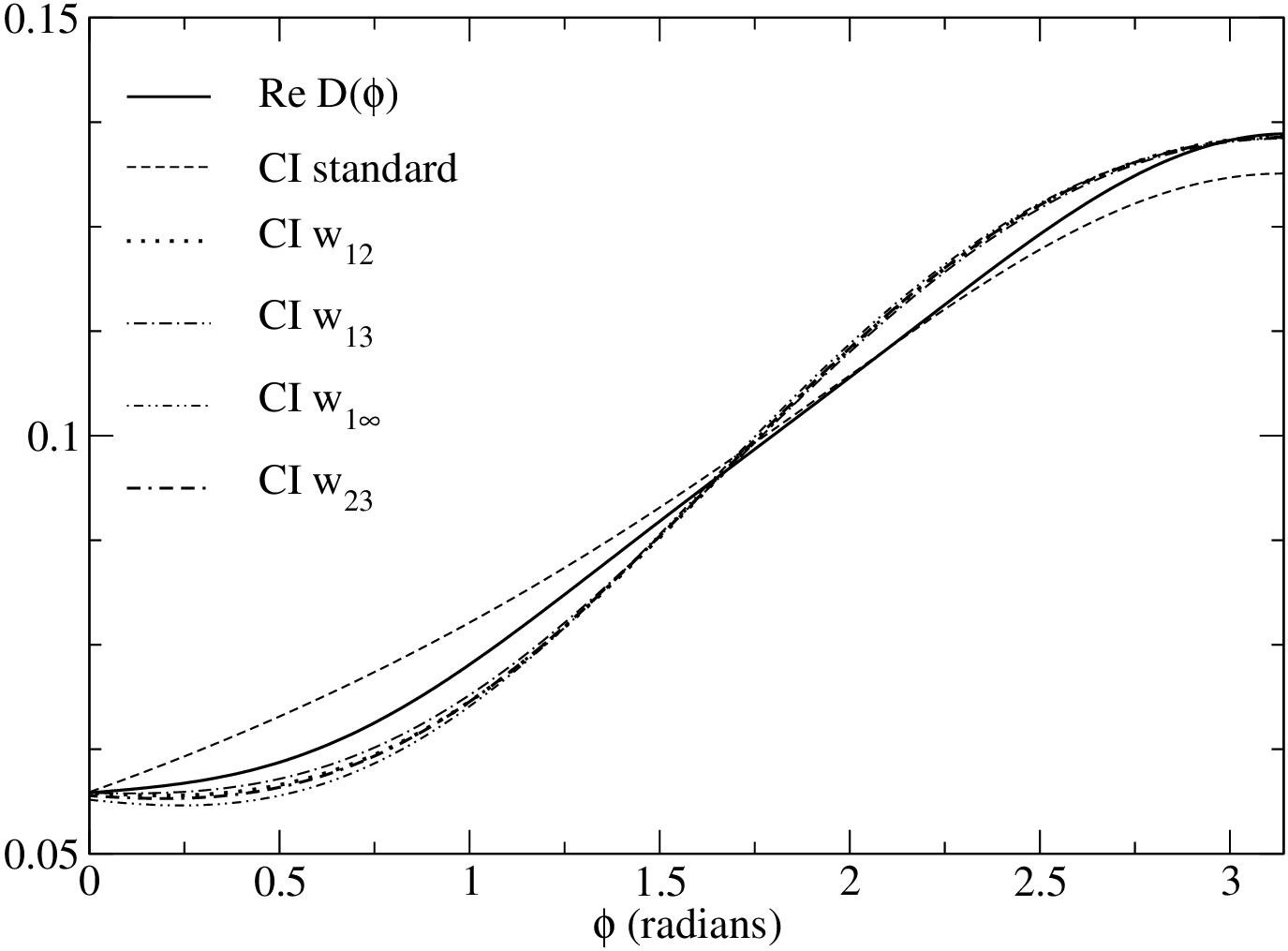}\hspace{1cm}
\includegraphics[width=6.cm]{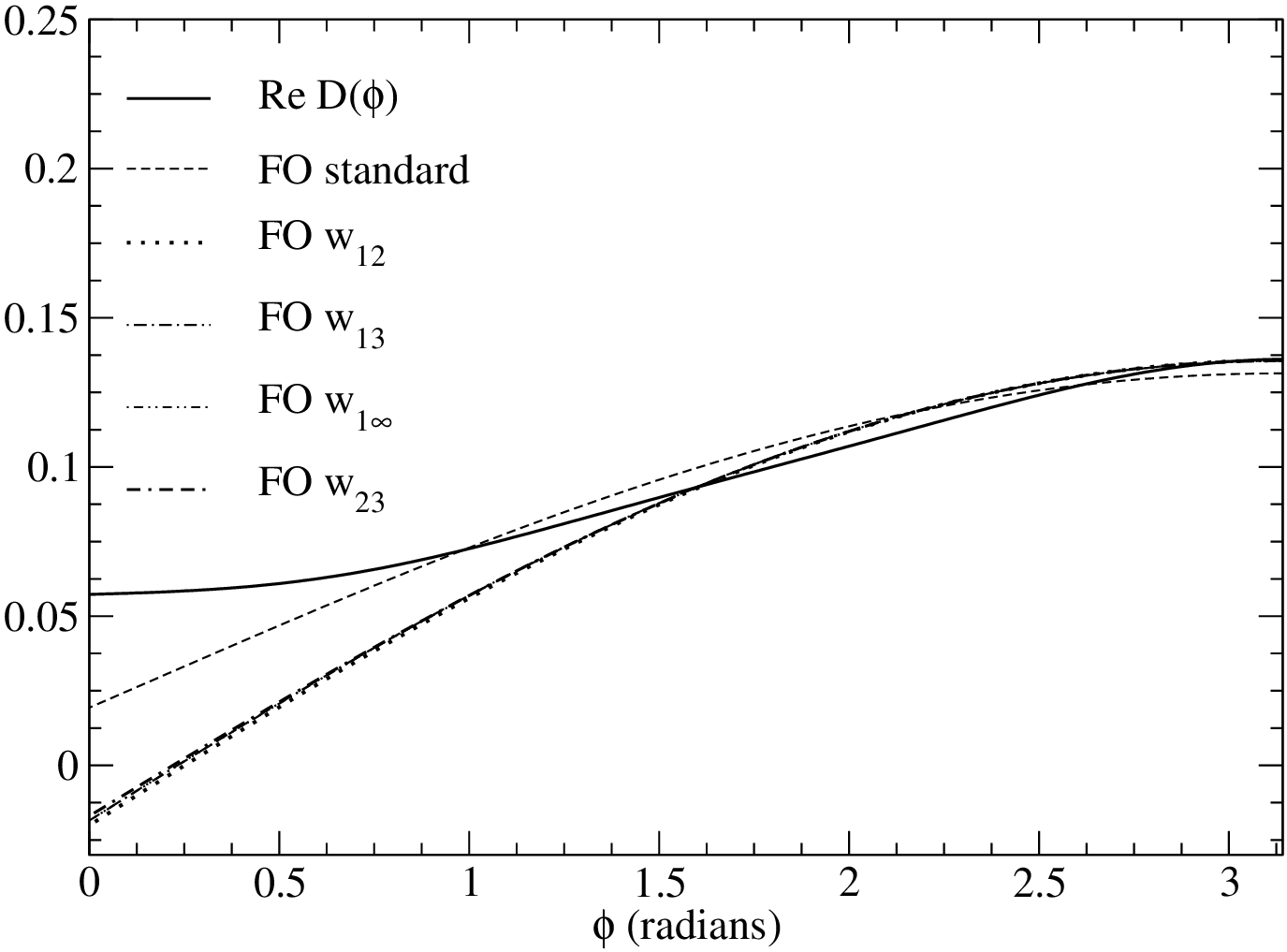}
\caption{\label{fig:DRalt5} Real part of the Adler function of the alternative model defined in (\ref{eq:altBBJ})-(\ref{eq:altdBJ1}), calculated along the circle $s=M_\tau^2 \exp(i\phi)$ for $\alpha_s(M_\tau^2)= 0.3156$, using the perturbative expansions with  $N=5$ terms. Left panel: CI expansions. Right panel: FO expansions. The exact function is represented by the solid line.}\end{center}\vspace{0.5cm}
\end{figure*}

\begin{figure*}[!ht]\begin{center}\includegraphics[width=6.cm]{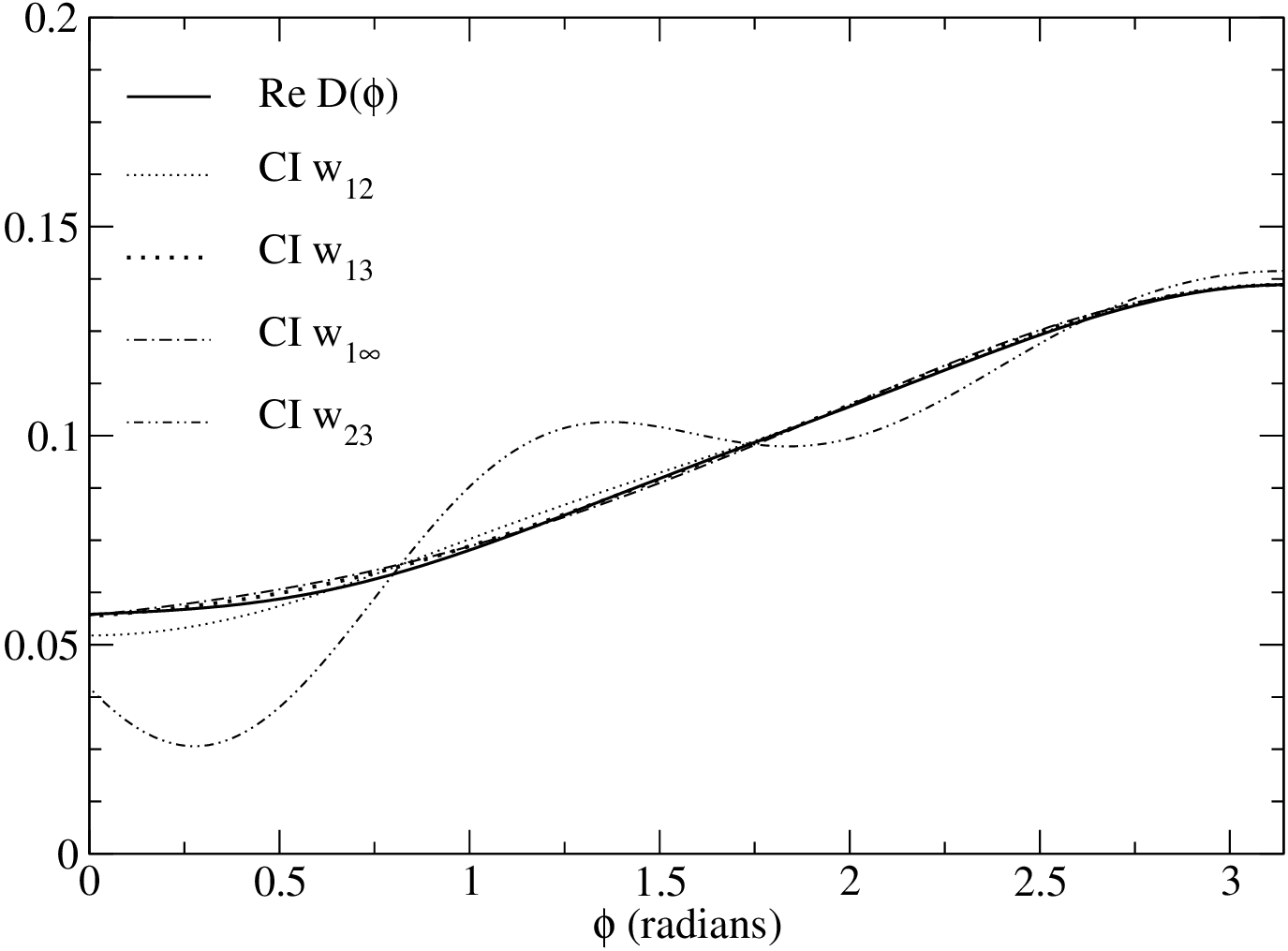}\hspace{1cm}\includegraphics[width=6.cm]{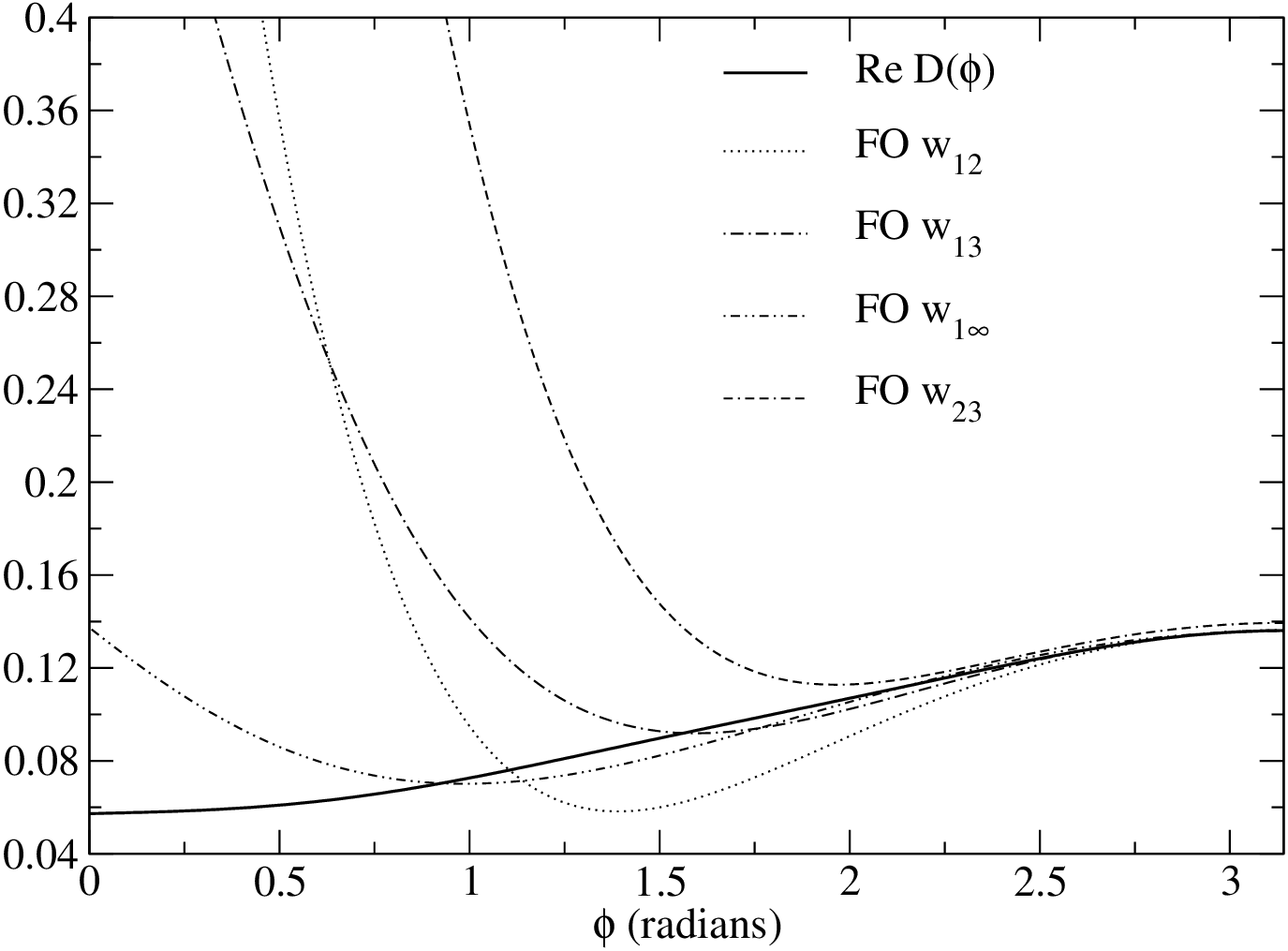}
\caption{\label{fig:DRalt18}  As in Fig. \ref{fig:DRalt5} for $N=18$.  The standard CI and FO expansions exhibit big oscillations and are not shown.}\end{center}\vspace{0.5cm}
\end{figure*}

\begin{table*}[!ht]
\caption{ The quantity $\delta^{(0)}$  for the model $B_{\rm BJ}$ proposed in \cite{BeJa} and specified in (\ref{eq:BBJ})-(\ref{eq:dBJ}), calculated for  
$\alpha_s(M_\tau^2)=0.34$ with the standard and modified  CI and FO expansions truncated at the order $N$. Exact value $\delta^{(0)} =0.2371$.} \vspace{0.1cm}
\label{table:1}
\renewcommand{\arraystretch}{1.1} 
\begin{tabular}{llllllllllr}\hline
$N$&CI st. &FO st.& CI $w_{12}$  &  FO $w_{12}$ &CI $w_{13}$& FO $w_{13}$ & CI $w_{1\infty}$& FO $w_{1\infty}$& CI $w_{23}$ &FO $w_{23}$\\\hline 
2& 0.1776& 0.1692& 0.1977& 0.2228& 0.2070& 0.2203& 0.1883& 
0.2524& 0.2123& 0.2099\\
3& 0.1898& 0.2026& 0.2009& 0.2460& 
0.2030& 0.2440& 0.1975& 0.2530& 0.2028& 0.2437
\\ 4& 0.1983& 
0.2200& 0.2263& 0.2463& 0.2194& 0.2460& 0.2288& 0.2465& 0.2206& 
0.2463\\ 5& 0.2022& 0.2288 & 0.2290 & 0.2440 & 0.2268 & 0.2423 & 
0.2310 & 0.2427 & 0.2292 & 0.2423 \\ 6& 0.2046 & 0.2328 & 0.2324 & 
0.2484 & 0.2306 & 0.2421 & 0.2321 & 0.2431 & 0.2319 & 0.2449 \\7& 
0.2046 & 0.2342 & 0.2339 & 0.2536 & 0.2331 & 0.2457 & 0.2333 & 
0.2454 & 0.2345 & 0.2502\\ 8& 0.2017 & 0.2353 & 0.2339  & 0.2505 & 
0.2343 & 0.2484 & 0.2341 & 0.2471 & 0.2347 & 0.2476 \\ 9 & 0.2004 & 
0.2367 & 0.2341 & 0.2431 & 0.2348 & 0.2457 & 0.2346 & 0.2465 & 
0.2347 & 0.2377 \\ 10 & 0.1842 & 0.2390 & 0.2351 & 0.2420 & 0.2348 & 
0.2394 & 0.2348 & 0.2436 & 0.2353 & 0.2337\\11 & 0.1962 & 0.2402 & 
0.2359 & 0.2406 & 0.2348 & 0.2352 & 0.2349 & 0.2399 & 0.2348 & 
0.2335\\12 & 0.1123 & 0.2436 & 0.2362 & 0.2298 & 0.2351 & 0.2349 & 
0.2349 & 0.2370 & 0.2374 & 0.2262\\ 13 & 0.2629 & 0.2408 & 0.2362 & 
0.2229 & 0.2355 & 0.2341 & 0.2349 & 0.2356 & 0.2348 & 0.2226\\ 14 & 
-0.2915 & 0.2575 & 0.2364 & 0.2242 & 0.2361 & 0.2303 & 0.2349 & 
0.2354 & 0.2395 & 0.2314\\ 15 & 1.1011 & 0.2170 & 0.2367 & 0.2173 & 
0.2366 & 0.2277 & 0.2350 & 0.2357 & 0.2356 & 0.2365\\ 16 & -3.362 & 
0.3818 & 0.2368 & 0.2102 & 0.2369 & 0.2305 & 0.2351 & 0.2360 & 
0.2343 & 0.2374\\17 & 9.5931 & -0.1881 & 0.2368 & 0.2176 & 0.2372 & 
0.2356 & 0.2352 & 0.2360 & 0.2533 & 0.2512\\ 18 & -31.52 & 2.144 & 
0.2368 & 0.2201 & 0.2373 & 0.2371 & 0.2354 & 0.2359 & 0.1926 & 
0.2665\\\hline \end{tabular}
\end{table*}

\begin{table*}[!htb]
\caption{The quantity $\delta^{(0)}$ for the modified model $B_{\rm alt}$ specified in (\ref{eq:altBBJ})-(\ref{eq:altdBJ1}), calculated for  
$\alpha_s(M_\tau^2)=0.34$  with the standard and modified  CI and FO expansions truncated at the order $N$. The rows for $N\le 5$ are identical to those in Table \ref{table:1}. Exact value $\delta^{(0)}=0.2102$.}\vspace{0.1cm}
\label{table:2}
\renewcommand{\tabcolsep}{0.5pc} 
\renewcommand{\arraystretch}{1.1} 
\begin{tabular}{llllllllllr}\hline
$N$&CI st. &FO st.& CI $w_{12}$  &  FO $w_{12}$ &CI $w_{13}$& FO $w_{13}$ & CI $w_{1\infty}$& FO $w_{1\infty}$& CI $w_{23}$ &FO $w_{23}$\\\hline 
6 & 0.2041 & 0.2318 & 0.2263 &  
0.2493 & 0.2271 & 0.2420 & 0.2284 & 0.2431 & 0.2260 & 0.2454 \\ 7 &  
0.2041 & 0.2290 & 0.2201 & 0.2628 & 0.2220 & 0.2481 & 0.2230 &  
0.2472 & 0.2174 & 0.2580 \\ 8 & 0.2023 & 0.2213 & 0.2202 & 0.2756 &  
0.2164 & 0.2595 & 0.2182 & 0.2541 & 0.2136 & 0.2734 \\ 9 & 0.2037 &  
0.2110 & 0.2175 & 0.2742 & 0.2143 & 0.2686 & 0.2154 & 0.2608 &  
0.2138 & 0.2706 \\ 10 & 0.1924 & 0.2032 & 0.2055 & 0.2709 & 0.2144 &  
0.2651 & 0.2146 & 0.2629 & 0.2115 & 0.2517 \\ 11 & 0.2124 & 0.2004 &  
0.1982 & 0.2905 & 0.2136 & 0.2504 & 0.2146 & 0.2578 & 0.2068 &  
0.2531 \\ 12 & 0.1412 & 0.2071 & 0.2007 & 0.3063 & 0.2111 & 0.2406 &  
0.2148 & 0.2468 & 0.2081 & 0.2627 \\ 13 & 0.3121 & 0.2117 & 0.2022 &  
0.2820 & 0.2086 & 0.2449 & 0.2149 & 0.2340 & 0.2060 & 0.2133 \\ 14 &  
-0.2105 & 0.2344  & 0.2001 & 0.2666 & 0.2074 & 0.2459 & 0.2146 &  
0.2239 & 0.2124 & 0.1338 \\ 15 & 1.2336 & 0.1934 & 0.2009 & 0.2865 &  
0.2079 & 0.2176 & 0.2142 & 0.2187 & 0.2087 & 0.1192 \\ 16 & -3.147 &  
0.3500 & 0.2044 & 0.2562 & 0.2091 & 0.1676 & 0.2136 & 0.2175 &  
0.2073 & 0.0930 \\ 17 & 9.948 & -0.2333 & 0.2059 & 0.1822 & 0.2102 &  
0.1355 & 0.2130 & 0.2175 & 0.2275 & -0.0415 \\ 18 & -30.94 & 2.084 &  
0.2058 & 0.1722 & 0.2107 & 0.1345 & 0.2124 & 0.2159 & 0.1617 &  
-0.1019 \\\hline\vspace{0.5cm} 
 \end{tabular}
 \end{table*}

\begin{figure*}[!ht]\begin{center}\includegraphics[width=7.cm]{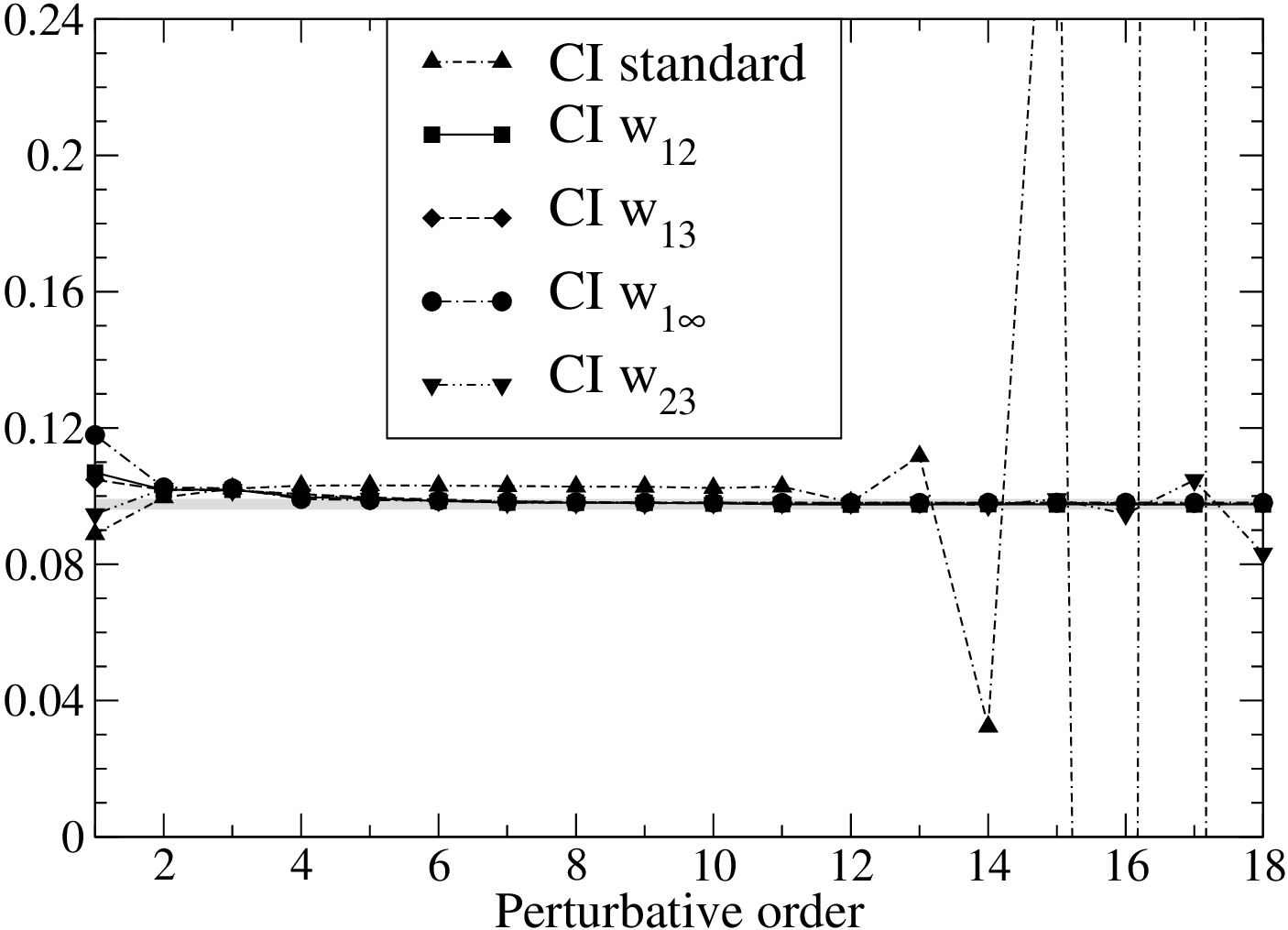}\hspace{1cm}
\includegraphics[width=7.cm]{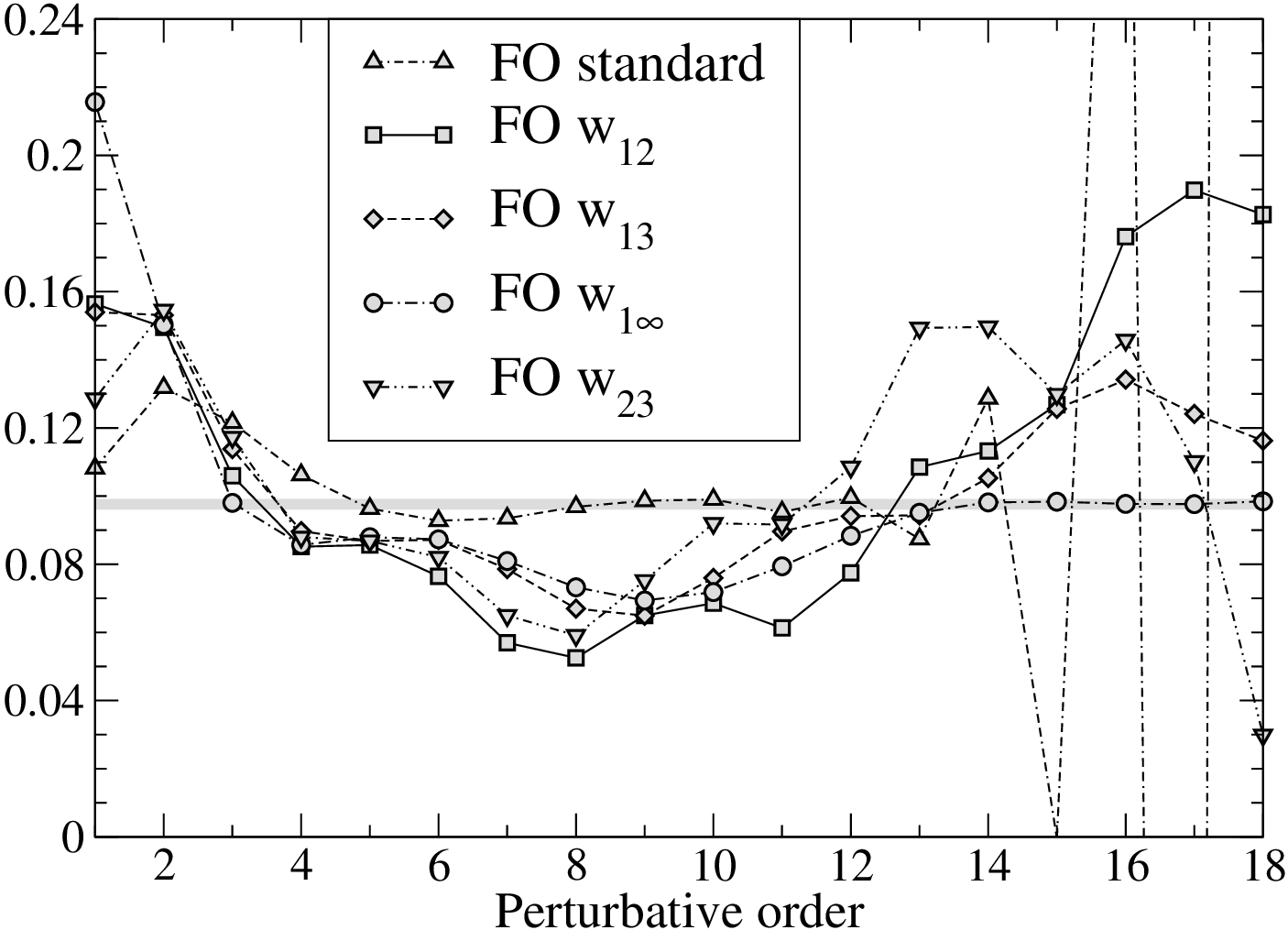}
\caption{\label{fig:MN5BJ} Moment $ M_5$ defined in (\ref{eq:Mk}) for the model \cite{BeJa}, calculated for $|s_0|=M_\tau^2$ and $\alpha_s(M_\tau^2)=0.34$, as a function of the perturbative order  $N$,  for the standard and the new expansions based on several conformal mappings $w_{jk}$ defined in (\ref{eq:wjk}). The grey horizontal line is the exact value.  Left panel: CI expansions. Right panel: FO expansions.}\end{center}\vspace{0.5cm}
\end{figure*}

\begin{figure*}[!ht]\begin{center}\includegraphics[width=7.cm]{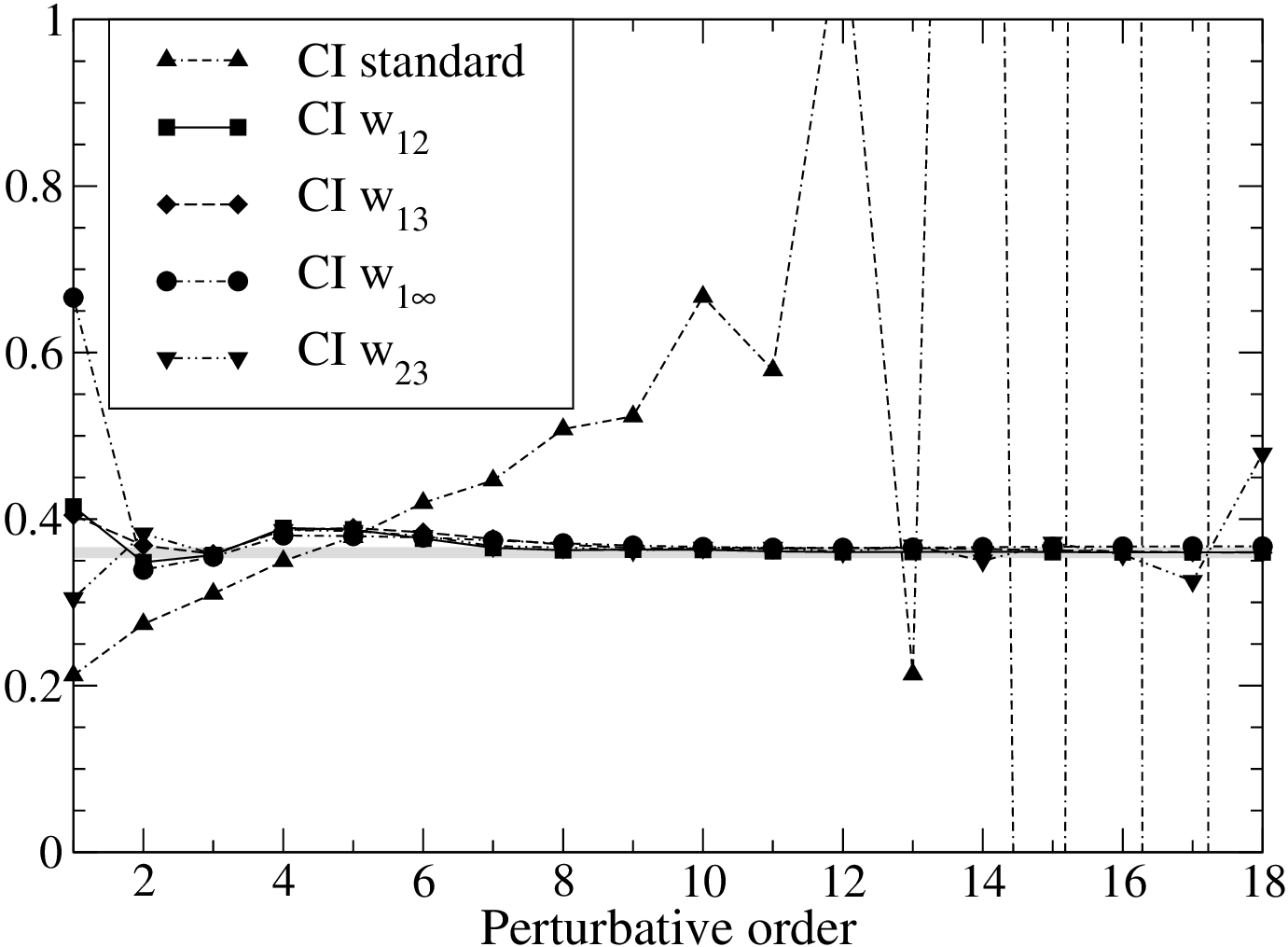}\hspace{1cm}\includegraphics[width=7.cm]{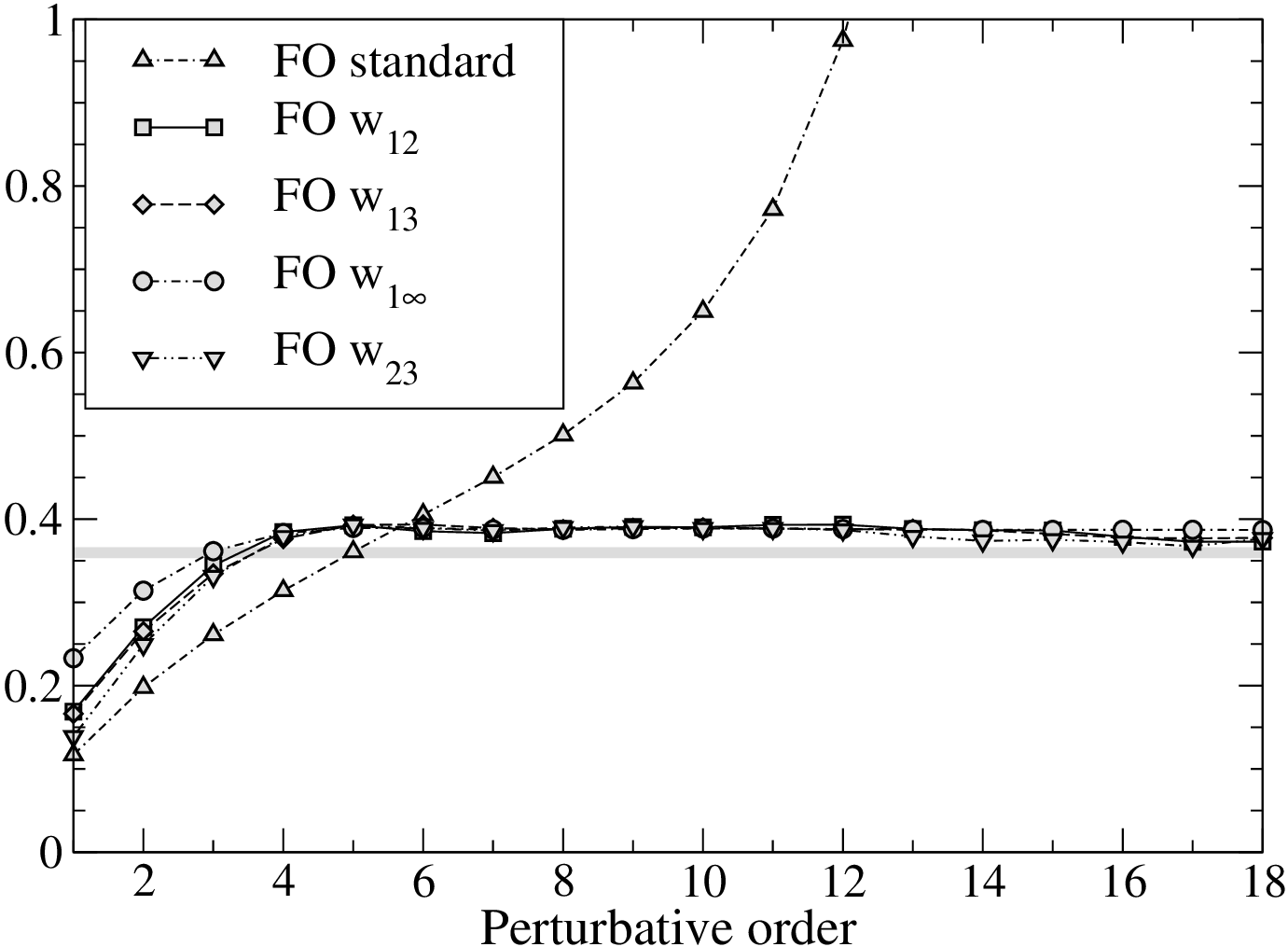}
\caption{\label{fig:MJ5BJ} Moment $\bar M_5$ defined in (\ref{eq:barMk}) for the model \cite{BeJa}, calculated for $|s_0|=M_\tau^2$ and  $\alpha_s(M_\tau^2)=0.34$, as a function of the perturbative order  $N$,  for the standard and the new expansions based on several conformal mappings. The grey horizontal line is the exact value. Left panel: CI expansions. Right panel: FO expansions. } \end{center} \vspace{0.5cm}
\end{figure*}

\section{Models}\label{sec:models}

For testing the convergence of the various expansions, we consider a class of  models of the type proposed in \cite{BeJa} ((but analyzed without using 
conformal mappings), which parametrize the Borel transform $B(u)$ and then recover the Adler function  by means of the PV prescription (\ref{eq:pv}).

 In the model proposed in \cite{BeJa}, the function $B(u)$ is expressed in terms of a few UV and IR renormalons 
\be\label{eq:BBJ}
B_{\rm BJ}(u)=B_1^{\rm UV}(u) +  B_2^{\rm IR}(u) + B_3^{\rm IR}(u) +d_0^{\rm PO} + d_1^{\rm PO} u,
\ee
 parametrized  as 
\be\label{eq:BIR}
B_p^{\rm IR}(u)= 
\frac{d_p^{\rm IR}}{(p-u)^{\gamma_p}}\,
\Big[\, 1 + \tilde b_1 (p-u) + \tilde b_2 (p-u)^2 +\ldots \,\Big], 
\ee
\be\label{eq:BUV}
B_p^{\rm UV}(u)=\frac{d_p^{\rm UV}}{(p+u)^{\bar\gamma_p}}\,
\Big[\, 1 + \bar b_1 (p+u) + \bar b_2 (p+u)^2 +\ldots \,\Big].
\ee
Most of the parameters are fixed using a renormalization-group analysis at four loops, the free parameters of the models being the residues $d_1^{\rm UV}, d_2^{\rm IR}$ and  $d_3^{\rm IR}$ of the first renormalons and the coeficients $d_0^{\rm PO}, d_1^{\rm PO}$ of the polynomial in (\ref{eq:BBJ}). They were fixed in  \cite{BeJa} by the requirement to reproduce the perturbative coefficients $K_n$ for $n\le 4$ from (\ref{eq:Kn}) and the estimate $K_5=283$, and are:
\begin{equation}
d_1^{\rm UV}=-\,1.56\times 10^{-2},~~~
d_2^{\rm IR}=3.16, ~~~
d_3^{\rm IR}=-13.5,\nn\\[-1mm]
\end{equation}
\begin{equation}\label{eq:dBJ}
d_0^{\rm PO}=0.781, ~~~
d_1^{\rm PO}=7.66\times 10^{-3}. 
\end{equation}
  We also investigated   alternative models, where we imposed a specific residue at $u=2$. In one such example, we kept the same expressions as in \cite{BeJa} for the first three singularities and  the same values of the  residues at $u=-1$ and $u=3$, while choosing a smaller residue at $u=2$, $d_2^{\rm IR}=1$.  The model must contain then three additional free parameters  in order to reproduce the first five $K_n$. Specifically, we introduced a quadratic term in the polynomial and two additional IR singularities, at $u=4$ and $u=5$. For convenience, the nature of these additional singularities, which is not known, was taken to be same as that of the $u=3$ singularity. Thus, we considered the alternative model:
\bea\label{eq:altBBJ}
B_{\rm alt}(u) &=&B_1^{\rm UV}(u) +  B_2^{\rm IR}(u) + B_3^{\rm IR}(u) + \frac{d_4^{\rm IR}}{(4-u)^{3.37}}\nn\\
&+&\frac{d_5^{\rm IR}}{(5-u)^{3.37}} + d_0^{\rm PO} + d_1^{\rm PO} u+ d_2^{\rm PO} u^3,
\eea
where, as  discussed above, we took as input
\be\label{eq:altdBJ}
d_1^{\rm UV}=-\,1.56\times 10^{-2},~~~
d_2^{\rm IR}=1, ~~~
d_3^{\rm IR}=-13.5,
\ee
and determined the remaining five parameters by matching the same coefficients $K_n$ for $n\le 5$: 
\be
d_0^{\rm PO}=3.2461, ~~~
d_1^{\rm PO}=1.3680, ~~~ d_2^{\rm PO}=0.2785,\nn\\[-3mm] 
\ee
\be\label{eq:altdBJ1}
d_4^{\rm IR}=1560.614, ~~~d_5^{\rm IR}=-1985.73.
\ee
 We emphasize that we consider these models only as a mathematical frame to test the convergence properties of the various expansions. The physical plausibility of one model or another \cite{BeJa,DeMa}  will not be discussed here.

In Figs. \ref{fig:DR5} and \ref{fig:DR18},  we show the real part of the Adler function for the  model \cite{BeJa}, calculated  along the circle $s=M_\tau^2 \exp(i\phi)$ with the standard and the new CI and FO expansions defined in (\ref{eq:Djk})-(\ref{eq:WnFOjk}), where the perturbative expansions were truncated at $N=5$ and $N=18$, respectively. To facilitate the comparison with previous works \cite{BeJa, CaFi2009}, we took $\alpha_s(M_\tau^2)=0.3156$ in this calculation. For $N=18$, the standard expansions exhibit big oscillations and are not shown.

The curves show that the new CI expansions based on  conformal mappings give a good approximation, which improves with increasing $N$, of the real part of $\wh D(s)$ along the whole circle (only  the mapping  $w_{23}$ shows signs of divergence for large $N$, as expected). This behavior is valid also at higher $N$ (we explored values up to $N=25$), for both the real and imaginary parts of the Adler function. 

\begin{figure*}[!ht]\begin{center}\includegraphics[width=7.cm]{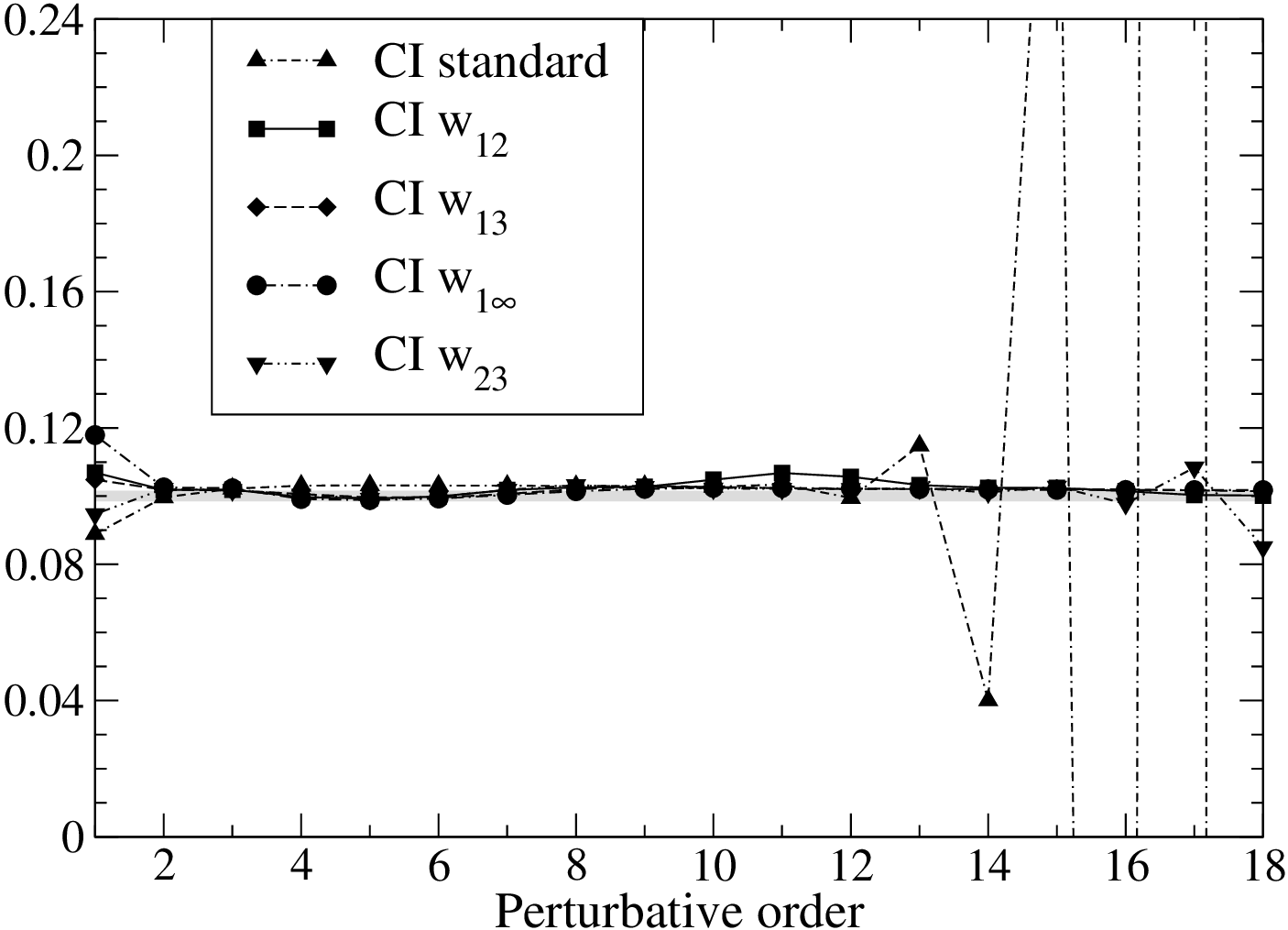}\hspace{1cm}
\includegraphics[width=7.cm]{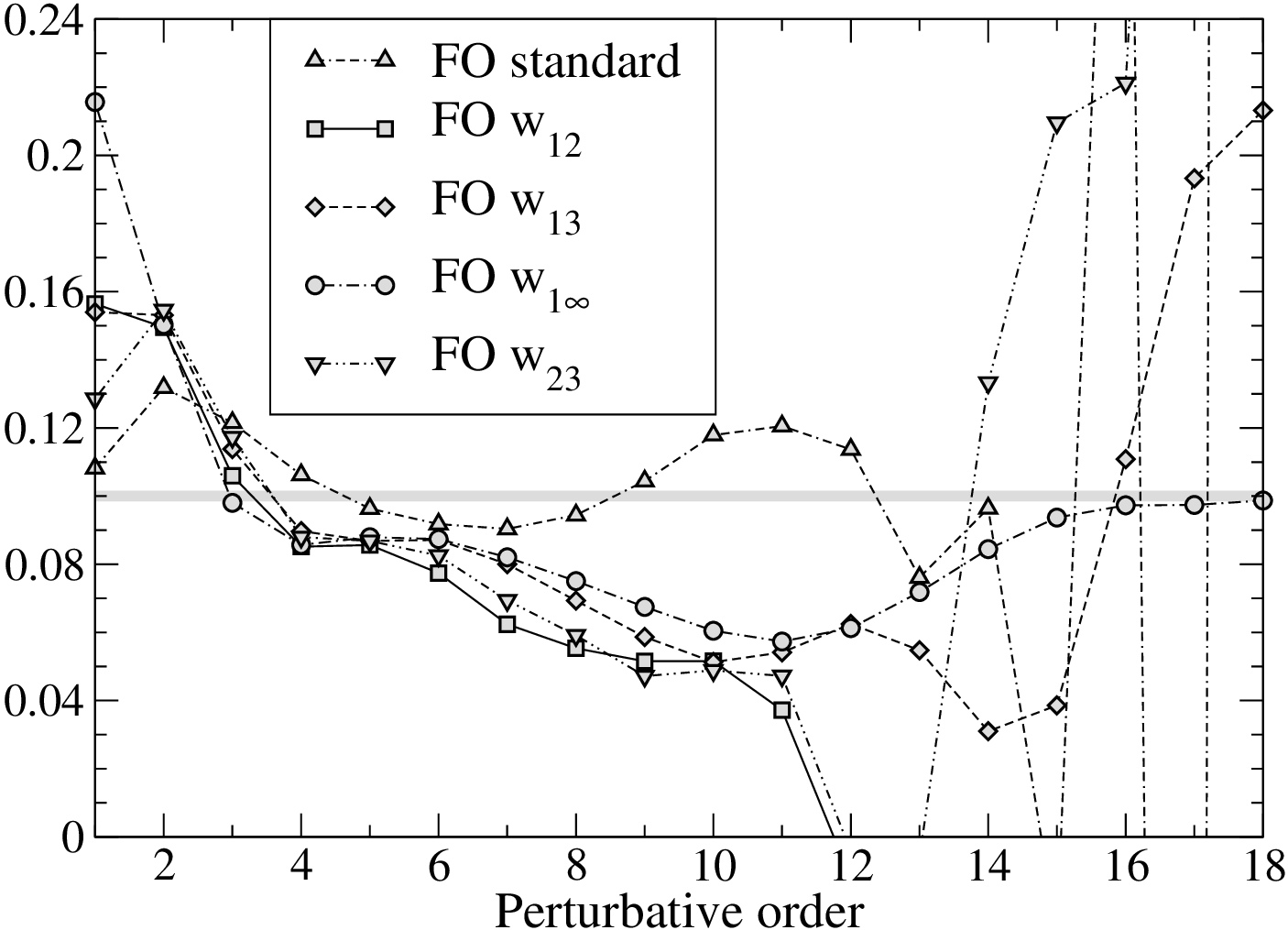}
\caption{\label{fig:MN5alt} Moment $M_5$ defined in (\ref{eq:Mk}) for the alternative model  (\ref{eq:altBBJ})-(\ref{eq:altdBJ1}), calculated for $|s_0|=M_\tau^2$ and $\alpha_s(M_\tau^2)=0.34$, as a function of the perturbative order  $N$  for the standard and the new expansions. The grey horizontal line is the exact value.  Left panel: CI expansions. Right panel: FO expansions.}\end{center}\vspace{0.5cm}
\end{figure*}

\begin{figure*}[!ht]\begin{center}\includegraphics[width=7.cm]{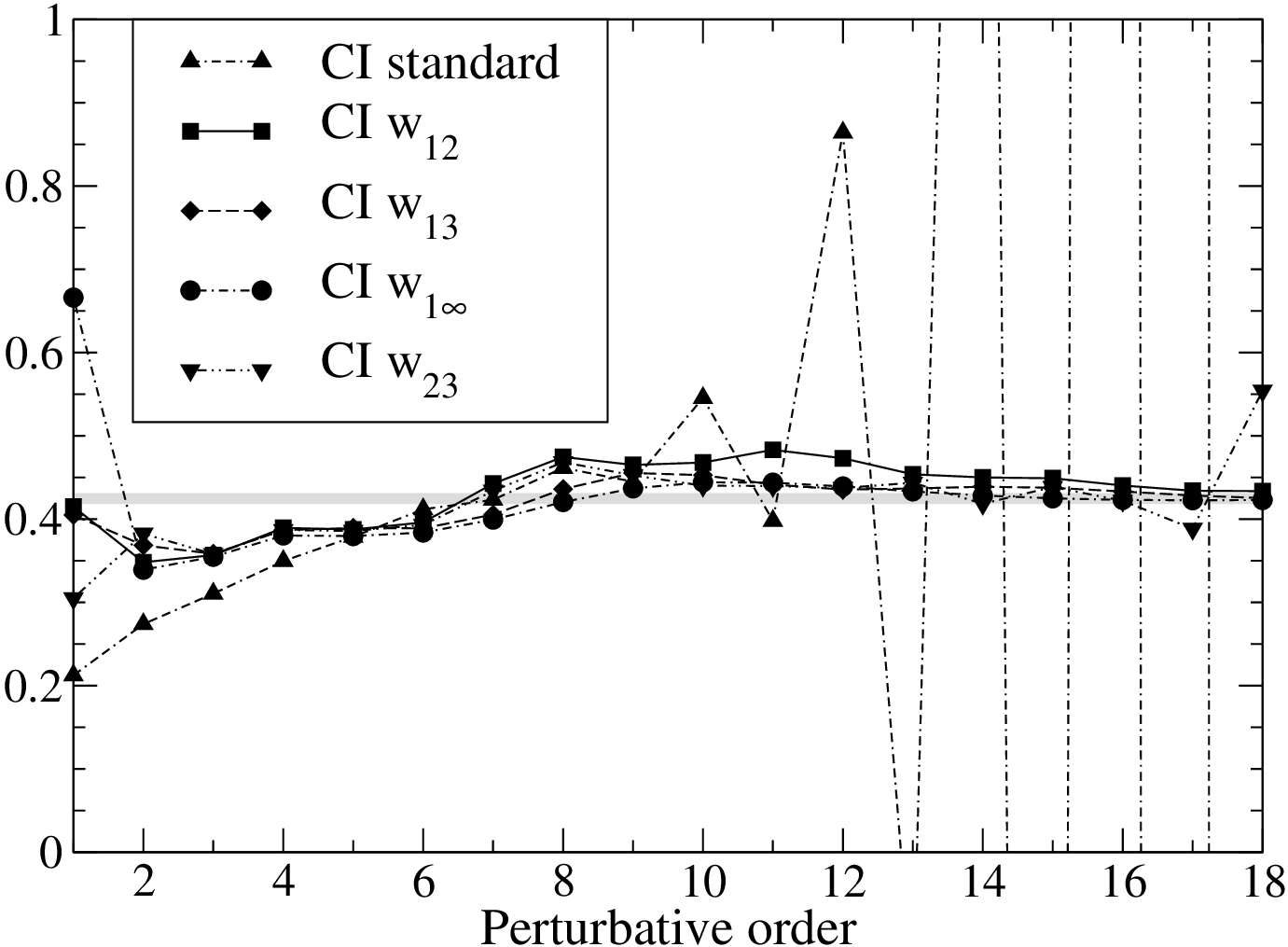}\hspace{1cm}
\includegraphics[width=7.cm]{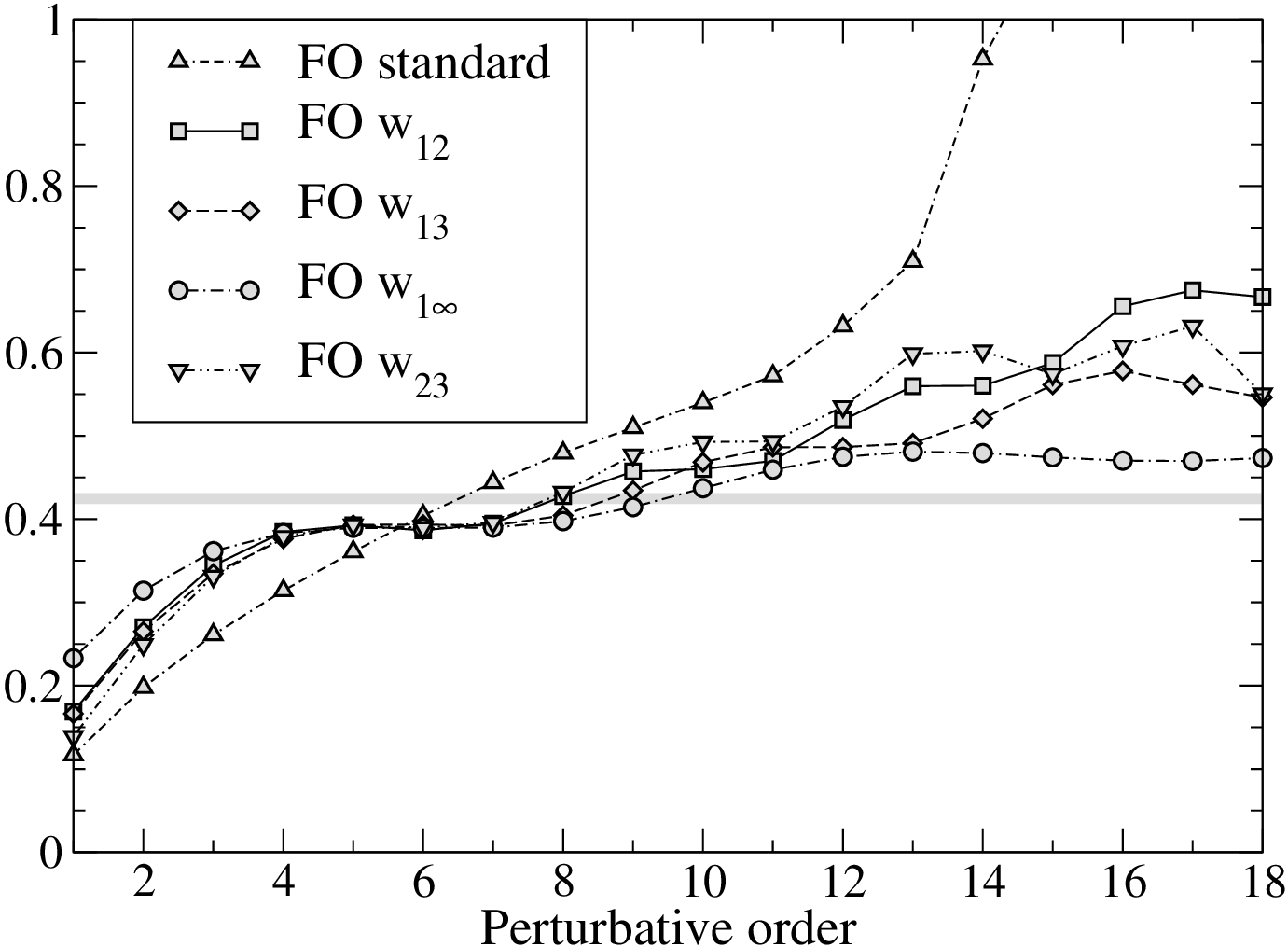}
\caption{\label{fig:MJ5alt}Moment $\bar{M}_5$ defined in (\ref{eq:barMk}) for the alternative model  (\ref{eq:altBBJ})-(\ref{eq:altdBJ1}), calculated for $|s_0|=M_\tau^2$ and $\alpha_s(M_\tau^2)=0.34$, as a function of the perturbative order  $N$  for the standard and the new expansions. The grey horizontal line is the exact value.  Left panel: CI expansions. Right panel: FO expansions.} \end{center}\vspace{0.5cm}
\end{figure*}

As concerns the FO expansions,  the description they provide is quite good for points close to the spacelike axis, $\phi=\pi$, but gradually deteriorates near the timelike axis, $\phi=0$.  To understand this behavior, we remark that the coupling $a_s(-s)$ is calculated along the circle as the exact solution of the RG equation in terms of $a_s(M_\tau^2)$, both in the "true" function (\ref{eq:pv}) and the CI expansion functions (\ref{eq:Wnjk}) (recall that $a_s=\alpha_s/\pi$). Therefore, the improvement of the series achieved by the conformal mappings  is clearly seen along the whole circle in the case of the CI expansions. On the other hand, the FO expansions are obtained by expanding  $a_s(-s)$  in powers of $a_s(M_\tau^2)$, according to (\ref{eq:astaylor}). As remarked in \cite{DiPi}, this expansion has a poor convergence near the timelike axis, due to the appearance of large imaginary logarithms in the coefficients. The curves in the right panels of Figs. \ref{fig:DR5} and \ref{fig:DR18}  show clearly the effect of the weak convergence of the additional series  (\ref{eq:astaylor}) involved in the definition of the FO expansions.  The detailed  behavior depends on the conformal mapping: for instance, the expansions based on the variables $w_{13}$ and $w_{1\infty}$ provide, for increasing $N$, a good approximation up to points rather close to  $\phi=0$, as shown in the right panel of Fig. \ref{fig:DR18}. 

In Figs. \ref{fig:DRalt5} and \ref{fig:DRalt18} we repeat the analysis for the alternative model (\ref{eq:altBBJ})-(\ref{eq:altdBJ1}). The perturbative curves in Figs. \ref{fig:DRalt5} coincide with those in Fig. \ref{fig:DR5}, since the first five perturbative coefficients of the two models coincide. On the other hand, the "true" function is slightly flatter in the second model, and  is better approximated by the standard  CI expansion with $N=5$ terms than it was the model shown in Fig.  \ref{fig:DR5}. By increasing $N$, the new CI expansions  based on conformal mappings again converge nicely towards the true function, as is seen in the left panel of Fig. \ref{fig:DRalt18}. The new FO expansions give a good approximation near the spacelike axis, but a poor description, even worse than for the previous model, near the  unitarity cut. 
The conclusion  is that, for both models,  the new CI expansions  give a precise approximation of $\wh D(s)$ along the whole circle, while the FO expansions give a description that deteriorates near the timelike axis. 

For the determination of $\alpha_s(M_\tau^2)$, the quantity of interest is the integral $\delta^{(0)}$ defined in (\ref{eq:delta0}). In  Table \ref{table:1}, we give the values of  $\delta^{(0)}$  for the model \cite{BeJa}, calculated   with the standard and the modified  CI and FO expansions, as a function of the perturbative order $N$. To facilitate the comparison with similar results reported in \cite{BeJa,CaFi2009} we took $\alpha_s(M_\tau^2)=0.34$.  As discussed in  \cite{BeJa}, at low $N$, the standard CI expansion gives values systematically lower than the true result, while the standard FO expansion gives a better approximation. As remarked already in \cite{CaFi2009}, for the same low values of  $N$, the difference between the results of the new CI and FO expansions is smaller than that of the standard ones. Moreover, as seen from Table \ref{table:1}, the difference decreases when passing from $N=4$ to  $N=5$, contrary to what happens with the standard expansions. At larger $N$, the new CI expansions approach the exact value  (deviations appear only for the expansion based on the conformal mapping $w_{23}$, for the reasons discussed in Sec. \ref{sec:expfct}). The new FO expansions give slightly worse  values, however,  the mappings $w_{13}$ and $w_{1\infty}$ lead to good  approximations at large $N$ also in the FO case.

 In Table \ref{table:2}, we present similar results for the alternative model (\ref{eq:altBBJ})-(\ref{eq:altdBJ1}).  By construction, the first five rows in Tables \ref{table:1} and \ref{table:2} are the same (but the "true" value is now different, $\delta^{(0)}=0.2102$ instead of  $\delta^{(0)}=0.2371$ in Table  \ref{table:1}). The CI expansions based on the mappings  $w_{12}$, $w_{13}$ and $w_{1\infty}$ approach at large $N$ the exact value also in this case.  In the FO case, the description is less precise and,  for the values of $N$ considered, only the expansion based on the mapping $w_{1\infty}$ exhibits  a good numerical convergence.

Finally, we illustrate the properties of the expansions  by calculating the perturbative part of the moments of the spectral function, defined in (\ref{eq:Mk}) and (\ref{eq:barMk}).  Detailed studies of the moments have been performed in Refs. \cite{Neubert,DiPi, MaYa, Pich_Manchester}, especially in connection with the power corrections.

 The approximation provided by various expansions results from the interplay between the behavior of the series and that of the integrand along the circle.  Both integrands in (\ref{eq:Mk}) and (\ref{eq:barMk}) vanish  on the timelike axis, but, while the second integrand suppresses the contribution of a region, which increases with $k$,  near the timelike axis, the first exhibits oscillations increasing with $k$ along the circle, and vanishes also on the spacelike axis for odd $k$.  From the behavior shown in Figs. \ref{fig:DR5}-\ref{fig:DRalt18},  we expect therefore a better approximation by the new FO expansions of the moments (\ref{eq:barMk}) compared to  (\ref{eq:Mk}). We recall that the new FO expansions give a good description of the Adler function near the spacelike axis, but the accuracy deteriorates near the timelike axis, due to the poor convergence of the expansion (\ref{eq:astaylor}). For the new CI expansions, a more or less comparable description at low orders, depending on the specific integrand,  and a very good convergence at high $N$, are foreseen. For the standard expansions, the results depend on the fortuitous cancellations of the contributions along the circle, as in the case of $\delta^{(0)}$.

This expectation is confirmed in Figs. \ref{fig:MN5BJ}-\ref{fig:MJ5alt}, where we show the moments $M_5$ and $\bar M_5$ calculated with the standard and the new expansions  for the model (\ref{eq:BBJ})-(\ref{eq:dBJ}) defined in \cite{BeJa}, and for the alternative model defined in  (\ref{eq:altBBJ})-(\ref{eq:altdBJ1}). We chose a rather high moment to see clearly the difference between the behavior of (\ref{eq:Mk}) and (\ref{eq:barMk}).  In all cases, we took $s_0=M_\tau^2$. 

For the first model, the new CI expansions give a very good  description of both moments, as shown in the left panels of Figs.  \ref{fig:MN5BJ}-\ref{fig:MJ5BJ}. The new FO expansions give a rather poor description of the moment $M_5$, but a very good approximation of the moment $\bar M_5$. The right panel of Fig. \ref{fig:MJ5BJ} shows that, for an integrand that strongly suppresses the region near the timelike axis, the new FO expansions provide a very good description. 
As for the standard expansions, at low orders they give a better approximation of the moment $M_5$, for which suitable cancellations of the terms along the circle occur.  At larger $N$, both standard expansions show large deviations from the true result. 

For the second model, Figs.  \ref{fig:MN5alt}-\ref{fig:MJ5alt} indicate a similar pattern, with  a slightly worse approximation given by the new expansions  at low orders. 
For both models,  the most impressive feature is the good description of the moments by the new CI expansions at large orders (only the expansion based on the variable $w_{23}$ exhibits small deviations at large $N$, as expected from the discussion in Sec. \ref{sec:expfct}). 

\section{Determination of $\alpha_s(M_\tau^2)$}\label{sec:alphas}
The above analysis demonstrated the good convergence properties of the new contour-improved (CI) perturbative series based on singularity softening and expansions of the Borel transform in powers of suitable conformal mappings. We apply now these expansions for a determination of  $\alpha_s(M_\tau^2)$ from the experimental rate of hadronic $\tau$ decays. We emphasize that our calculation is not based on the models discussed in the previous section, but relies only on the known coefficients $K_n$ given in (\ref{eq:Kn}) and a very conservative choice \cite{BeJa, Beneke_Muenchen}  for the next coefficient, $K_5=283\pm 283$.   For the running of the coupling, we use the calculated perturbative coefficients $\beta_j$ from (\ref{eq:betaj}), an assumption about the next coefficient $\beta_4$ being considered only for the assessment of the errors.

The standard determination of $\alpha_s(M_\tau^2)$ from hadronic $\tau$-decays requires the theoretical calculation of the integral
defined in (\ref{eq:delta0}), using the perturbative expansion of the Adler function. On the other hand, the quantity $\delta^{(0)}$ can be determined  with great precision from (\ref{eq:RtauVA}).
The recent determination $R_{\tau, V+A}=3.4771\pm 0.0084$ \cite{HFAG} 
leads to the  updated phenomenological value \cite{Beneke_Muenchen} 
\be\label{eq:input}
\delta^{(0)}_{\rm phen}=0.2037 \pm 0.0040_{\rm exp} \pm 0.0037_{\rm PC},
\ee
where the first error is experimental and the second accounts for the power corrections.

Using this input, the values of  $\alpha_s(M_\tau^2)$ obtained with the  new CI expansions defined in (\ref{eq:Djk})-(\ref{eq:Wnjk}), with the expansion functions ${\cal W}_n^{12}$,  ${\cal W}_n^{13}$,  ${\cal W}_n^{1\infty}$ and  ${\cal W}_n^{23}$, respectively, are:
\begin{eqnarray}\label{eq:alphas}
&&\hspace{-0.45cm} 0.3195  \pm  0.0034_{\rm exp}  \pm 0.0031_{\rm PC} ~^{+0.0246 }_{- 0.0137}(K_5)~^{+ 0.0018}_{-0.0019 }{(\rm scale)},\nonumber\\
&&\hspace{-0.45cm} 0.3208\pm 0.0035_{\rm exp} \pm  0.0032_{\rm PC} ~^{+ 0.0131}_{-0.0093 }(K_5)~^{+ 0.0024}_{-0.0088 }{(\rm scale)},    \nonumber\\
&&\hspace{-0.45cm} 0.3182 \pm 0.0033_{\rm exp} \pm 0.0031_{\rm PC} ~^{+0.0172 }_{-0.0111}(K_5)~^{+ 0.0025}_{-0.0088 }{(\rm scale)},    \nonumber\\
&&\hspace{-0.45cm} 0.3193 \pm 0.0034_{\rm exp} \pm 0.0031_{\rm PC}  ~^{+ 0.0182}_{- 0.0115 }(K_5)~^{+ 0.0023}_{-0.0063 }{(\rm scale)}. \nonumber\\
\end{eqnarray}
 The first two errors are produced by the uncertainties of $\delta^{(0)}_{\rm phen}$ given in (\ref{eq:input}), the third one is obtained by varying the coefficient $K_5$ in the conservative range mentioned above, and the last error accounts for the variation of the scale as $ \xi M_\tau^2$, with  $\xi$  in the range $0.5-1.5$  \cite{Pich_Manchester}. 

The largest errors in (\ref{eq:alphas}) are  produced by the uncertainty in the coefficient $K_5$. 
To understand this result, we remark that the series  (\ref{eq:Djk})-(\ref{eq:Wnjk}), when reexpanded in powers of $\alpha_s$, generate an infinite number of higher-order terms \cite{CaFi1,CaFi2009}. In particular, the representations based on the expansion functions ${\cal W}_n^{12}$,  ${\cal W}_n^{13}$,  ${\cal W}_n^{1\infty}$ and  ${\cal W}_n^{23}$, truncated after four terms ({\em i.e.} neglecting the fifth term proportional to $K_5$), lead to coefficients $K_5$ equal to 256, 161, 256, and 179, respectively. If a very different value, like  $K_5=0$ or $K_5=566$, is imposed, the expansions can match the same $\delta^{(0)}_{\rm phen}$ only with the price of a much larger/smaller coupling, respectively.  In fact, if $K_5$  is assumed to be negative and large,  the coupling should be so large that the calculation becomes unreliable, and no solution  $\alpha_s(M_\tau^2)$ exists at all.

By taking the average  of the values in (\ref{eq:alphas}), we obtain
\begin{multline}\label{eq:aver}
\alpha_s(M_\tau^2)= 0.3195  \pm  0.0034_{\rm exp}  \pm 0.0031_{\rm PC} ~^{+0.0182}_{- 0.0114}(K_5)~\\^{+ 0.0018}_{-0.0019 }{(\rm scale)} \pm 0.0005_{\beta_4},  
\end{multline}
where we added an uncertainty to account for the truncation of the $\beta$-function (obtained by including a further term based on a geometrical growth, $\beta_4=\pm \beta_3^2/\beta_2$ \cite{Davier2008, Pich_Manchester}). We emphasize that the errors quoted in (\ref{eq:aver}) were obtained  as simple averages of the individual errors given in (\ref{eq:alphas}).  Much lower uncertainties would be obtained if standard statistical  procedures for independent determinations (for instance, Eqs. (14) and (15) of \cite{Bethke})   were applied. In practice, although the values given in (\ref{eq:alphas}) may be considered independent theoretical determinations, we prefer the conservative errors given in (\ref{eq:aver}), which avoid any bias. The remarkable consistency of the  theoretical determinations (\ref{eq:alphas}) is nevertheless a strong  argument in favor of our predictions. 

By combining in quadrature the errors given in (\ref{eq:aver}), we finally obtain 
\be\label{eq:aver1}
\alpha_s(M_\tau^2)= 0.3195~^{+0.0189}_{- 0.0138}.
\ee
The central value in (\ref{eq:aver1}) coincides practically with our previous determination \cite{CaFi2009}, $\alpha_s(M_\tau^2)= 0.320 \pm 0.011$, obtained with the optimal mapping $w_{12}$ and the slightly different value $\delta^{(0)}_{\rm phen}=0.2052 \pm 0.0050$ from \cite{BeJa}. The smaller error quoted in \cite{CaFi2009} is due mainly to a smaller range,  $K_5=283\pm 142$, adopted there for the coefficient $K_5$.
 
 We note that for the same input,  the standard CI expansion to 5-loops leads to
$\alpha_s(M_\tau^2)= 0.3419 \pm  0.012$, while the standard FO expansion gives $\alpha_s(M_\tau^2)= 0.3199^{+0.0118}_{-0.0074}$ \cite{Beneke_Muenchen}.   The smaller errors are mainly due to the fact that the standard expansions are less sensitive to the variation of  $K_5$.  However, these expansions have the behavior expected for an asymptotic series, approaching the expanded function up to a certain order $N$, and starting to oscillate violently afterwards. For some expanded functions,  the minimal error reached  before the onset of oscillations may be rather small, but in other cases, the standard expansion never describes the function with sufficient accuracy. Therefore, the uncertainty of  $K_5$ can generate only a part of the truncation error, an additional term  being necessary in order to account for the divergent pattern (this term may be taken, for instance, as the difference of about 0.022 between the predictions of the standard CI and FO expansions to 5-loops).

\section{Summary and conclusions}\label{sec:disc}

In this paper, we investigated a new class of expansions of the Adler function in perturbative QCD and applied them to a determination of $\alpha_s$ from hadronic $\tau$ decays. Our work extends previous studies reported in \cite{CaFi1,CaFi2, CaFi3, CaFi2009, CaFi_Manchester}. 

As remarked in \cite{ZJ1}, if a series is divergent and the expansion parameter is not very small, a summation of the perturbative expansion  is indispensable. The definition of the expansion functions investigated in this work exploits the information available on the large-order behavior of the perturbative series, together with  mathematical results on accelerating the series convergence by conformal mappings \cite{CiFi}. These techniques are suitable for the Borel plane, where an analyticity domain around the origin exists.  An important feature \cite{CaFi2} is that the expansion functions share the singularity of the expanded function at the origin $\alpha_s=0$ of the coupling plane. 

In the present study, we focused on the procedure of "singularity softening," by which the strong leading singularities in the Borel plane are turned into milder singularities, where the function vanishes instead of exploding. In practice, this is achieved by expanding  the product of $B(u)$ with suitable factors that vanish at the points $u=-1$ and $u=2$.  The procedure is possible due to the exact results available on the nature of the leading singularities \cite{Muell,BBK,BeJa}.  Since the effect of a mild singularity is expected to occur only at higher orders in a power series, one can choose as expansion variable a conformal mapping that accounts only for the further singularities of the Borel transform.  Extensive numerical studies showed that it is convenient to take the factors that multiply the function  $B(u)$ (which in principle are arbitrary) as simple expressions of the same variable that is used for expanding the product. 

In Sec. \ref{sec:models}, we investigated in detail the properties of the expansions  defined in Sec. \ref{sec:expfct} by using two specific models for the Adler function. As already mentioned, we consider these models only as a mathematical frame for testing the convergence properties of the various expansions and make no assumption about their physical plausibility (actually, there is no consensus on this subject in the literature). In all cases, we obtain a good convergence of the new contour-improved (CI)   expansions defined in (\ref{eq:Djk})-(\ref{eq:Wnjk}), for the choices of $j$ and $k$ adopted in this work.  The alternative fixed-order (FO) expansions  (\ref{eq:DFOjk})-(\ref{eq:WnFOjk}) converge near the spacelike axis, but provide a worse approximation near the timelike axis, since they also involve   the expansion (\ref{eq:astaylor}), which converges slowly in this region.  

Suitable cancellations between the two terms in  the expansion (\ref{eq:Dpert})  make the standard FO expansions more suitable  for calculating integrals like (\ref{eq:delta0}) for some models. Such fortuitous cancellations are not expected to occur in the case of the new FO expansions, where the approximation  is quite good near the spacelike axis and gradually deteriorates for points closer to the timelike axis. Thus,  the new CI expansions considered in the present work have a more solid theoretical basis than the new FO expansions.  

In Sec. \ref{sec:alphas}, we present a determination of $\alpha_s(M_\tau^2)$ based on the new CI expansions  defined in Eqs. (\ref{eq:Djk})-(\ref{eq:Wnjk}).  We emphasize that in our analysis we do not rely on models  and make no assumption about the strength of the leading singularities in the Borel plane. The predictions of the various expansions, reported in (\ref{eq:alphas}), exhibit a remarkable consistency among each other. 

Our final prediction  (\ref{eq:aver1}), obtained by averaging the individual values  (\ref{eq:alphas}), is very close to that of the standard FO expansion, while the standard CI expansion gives a value larger by about 0.022. It is important to understand the origin of this result. In our opinion, it is related to the consistent treatment of the running of the coupling and the Adler function coefficients in the standard FO expansion and the new CI one.

 As discussed in \cite{BeJa}, the standard FO expansion  is suitable for  models like the ansatz (\ref{eq:BBJ})-(\ref{eq:dBJ}), where the residues of the dominant renormalons are fixed in a natural way from the first coefficients $K_n$.  On the other hand, as noticed in Sec. \ref{sec:models}, this expansion is not so efficient for more artificial models like that presented in Eqs. (\ref{eq:altBBJ})-(\ref{eq:altdBJ1}),   where the strength of the  first IR renormalon is forced by hand to a lower value. So, the standard FO expansion  seems more suited than the standard CI one for describing functions with a natural pattern of leading singularities. Suitable compensations of the two terms of the same order in (\ref{eq:Dpert}) play an important role in this description.
 As noticed in \cite{Beneke_Muenchen}, these cancellations are destroyed in the standard CI expansion (\ref{eq:DpertCI}), which sums the running coupling terms, but drops the Adler function coefficients $K_n$ in higher orders.

 It is precisely this deficiency that is corrected by the new CI expansions, which sum also the  Adler function coefficients, by properly  implementing the singular behavior near the leading renormalons (with no assumptions about their strength) and expanding in powers of a conformal mapping.  So, the new CI expansions sum  both the running coupling terms and the expansion of the Adler function, while in the standard FO expansion, the fixed-order option is made for both expansions. This symmetric treatment explains why their predictions are similar, at least for truncation orders $N=4$ or $N=5$ of interest at present. At higher $N$, while both the standard FO and CI expansions start to diverge, the new CI expansions show an impressive convergence for all types of expanded functions. 

As we already discussed,  the largest errors in (\ref{eq:alphas}) are due to the conservative range $K_5=283 \pm 283$ adopted for the 5-loop coefficient $K_5$.  Expressed in other words, the new expansions  appear to exclude both very large and very small (or negative) values of $K_5$, which would require unusual values  of $\alpha_s(M_\tau^2)$ to reproduce the input (\ref{eq:input}). For the standard expansions, the error related to $K_5$ is much smaller, but it cannot fully account for the asymptotic character of the series, which may start to oscillate at a certain $N$  without approaching the expanded function with a sufficient accuracy.  

The value given in  (\ref{eq:aver1}) represents our best determination,  obtained as the average of the determinations (\ref{eq:alphas}) with a very conservative treatment of the uncertainties.  Our analysis shows that for increasing the precision   of  $\alpha_s(M_\tau^2)$ determination with the new expansions a more precise knowledge of the 5-loop coefficient  $K_5$ is crucial.

\vspace{0.cm}
\subsection*{Acknowledgements} We thank M. Beneke, M. Jamin, A. Pich and I. Vrko\u c for very useful discussions and suggestions during the work on this paper. I.C. acknowledges support from CNCSIS in the program Idei, Contract No.464/2009. J.F. acknowledges support from project No. LA08015 of the Ministry of Education and project No. AVO-Z10100502 of the Academy of Sciences of the Czech Republic. 

\vskip0.cm

\appendix

\section{Proof of lemmas given in Sec. \ref{sec:confmap}}

{\large $\bullet$}  {\em Proof of Lemma 1:} We  remind the reader that Lemma 1 was stated and proved in Ref. \cite{CiFi}; we recall it for completeness here. 

 Let us define
\begin{equation} \label{eq:f}
f(z_2) = \tilde z_1(\tilde z_2^{[-1]}(z_2))   
\end{equation}
for $z_2 \in {\cal K}_2$, where $\tilde z_2^{[-1]}$ is the inverse to $\tilde z_2$, which
exists since $\tilde z_2(u)$ is a conformal mapping.

The function $f(z_2)$ is holomorphic on the unit disk $ {\cal K}_2$ of the $z_2$-plane and maps this disk   into the unit disk $ {\cal K}_1$ of the $z_1$-plane, {\it i.e.}  $|f(z_2)| \leq 1$.
Moreover, since $\tilde z_1(Q)=0$ and  $\tilde z_2(Q)=0$  by assumption, it follows that $f(0)=0$. 

We now apply Schwarz's lemma, which states that if a function $F(z)$ is holomorphic on the disk $|z|<1$ and satisfies the
conditions $F(0)=0$ and $|F(z)|<1$ for $|z|<1$, then 
\begin{equation} \label{eq:Sch}
|F(z)| \leq |z|    
\end{equation}
everywhere in $|z|<1$. Besides, if the equality sign occurs in (\ref{eq:Sch}) at  least at one
interior point, then it takes place everywhere and $F(z)$ has the form
$F(z)=z \exp(i \alpha)$ with $\alpha$ real.

Applying Schwarz's lemma to the function $f$ defined in (\ref{eq:f}), we
have $|f(z_2)| \leq |z_2|$  for $z_2 \in {\cal K}_2$. Using the definition (\ref{eq:f}) and  the obvious relation
$\tilde z_2^{[-1]}(z_2)=u$ for $u\in {\cal D}_2$, we obtain 
\be\label{eq:proof1}
|\tilde z_1(u)| \leq |\tilde z_2(u)|,\quad u\in {\cal D}_2.
\ee
Ignoring the mappings that reduce to mere rotations according to Schwarz's lemma,
we are left with a sharp inequality in (\ref{eq:proof1}),
\be\label{eq:proof2}
|\tilde z_1(u)| < |\tilde z_2(u)|,\quad u\in {\cal D}_2, \quad u\ne Q,\ee
which proves  Lemma 1.

\vspace{0.3cm}
{\large $\bullet$} {\em Proof of Lemma 2:}  The relations (\ref{eq:limsup}) 
 imply that the coefficients $|c_{n,j}|$ can, for large enough $n$, be represented in the form
\begin{equation}\label{eq:cn}
|c_{n,j}|= e^{g_j(n)},\quad \quad j=1,2,
\end{equation}
where  $g_j(n)$ are real-valued functions, subject to the conditions $ \lim_{n \to \infty}g_j(n)/n = 0$, $j=1,2$.
Then, the ratio defined in (\ref{eq:rate}) can be written as
\be\label{eq:Rn}
 {\cal R}_n(u) = e^{g(n)}\times (\rho(u))^n,
\ee
where
\be\label{eq:gnrho}
g(n)=g_1(n)-g_2(n), \quad\quad \rho(u) =|\tilde z_1(u)/\tilde z_2(u)|. 
\ee
Taking the logarithm of (\ref{eq:Rn}), one obtains, for large $n$,  the inequality
\be \label{eq:lnr}
\ln {\cal R}_n(u) = n\left[\frac{g(n)}{n} + \ln \rho(u)\right] <0,
\ee
since from (\ref{eq:gnrho}) it follows that $\lim_{n \to \infty} g(n)/n =0$,  while 
$\rho(u)<1$ for all $u\in  {\cal D}_2$, $u\ne Q$, according to Lemma 1. This implies (\ref{eq:rate}), proving Lemma 2.


\end{document}